\documentclass[aps,prd,onecolumn,showpacs,superscriptaddress,preprintnumbers,floatfix,reprint]{revtex4-1}
\usepackage{graphicx}
\usepackage{amsmath}
\usepackage{bm}
\usepackage{yhmath}
\usepackage[colorlinks,citecolor=blue,urlcolor=blue,linkcolor=blue]{hyperref}
\usepackage{color}
\usepackage{epstopdf}
\usepackage{cases}
\usepackage{subfigure}
\usepackage{dcolumn,booktabs,bm}
\usepackage{slashed}
\usepackage{amsfonts,amssymb,stmaryrd,latexsym,amsmath}
\usepackage{textcomp}
\usepackage{multirow}
\usepackage{cancel}
\usepackage{array}

\newcommand{\etal}{\textit{et al}. }

\allowdisplaybreaks

\begin{document}

\title{The hidden charm pentaquark states and $\Sigma_c\bar{D}^{(*)}$ interaction in chiral perturbation theory}

\author{Lu Meng}\email{lmeng@pku.edu.cn}
\affiliation{School of Physics and State Key Laboratory of Nuclear
Physics and Technology, Peking University, Beijing 100871, China}

\author{Bo Wang}\email{bo-wang@pku.edu.cn}
\affiliation{School of Physics and State Key Laboratory of Nuclear
Physics and Technology, Peking University, Beijing 100871, China}
\affiliation{Center of High Energy Physics, Peking University,
Beijing 100871, China}

\author{Guang-Juan Wang}\email{wgj@pku.edu.cn}
\affiliation{School of Physics and State Key Laboratory of Nuclear
Physics and Technology, Peking University, Beijing 100871, China}
\affiliation{Center of High Energy Physics, Peking University,
Beijing 100871, China}

\author{Shi-Lin Zhu}\email{zhusl@pku.edu.cn}
\affiliation{School of Physics and State Key Laboratory of Nuclear
Physics and Technology, Peking University, Beijing 100871, China}
\affiliation{Center of High Energy Physics, Peking University,
Beijing 100871, China} \affiliation{Collaborative Innovation Center
of Quantum Matter, Beijing 100871, China}

\begin{abstract}
In this work, we employ the heavy hadron chiral perturbation theory
(HHChPT) to calculate the $\Sigma_c\bar{D}^{(*)}$ potentials to the
next-to-leading order. The contact, the one-pion exchange and the
two-pion exchange interactions are included. Besides, the mass
splittings between the heavy quark spin symmetry (HQSS) multiplets
are kept in calculations. Our result shows that neglecting the heavy
quark symmetry (HQS) violation effect may be misleading to predict
the potentials between the charmed hadrons. We perform numerical
analysis with three scenarios. In the first scenario, we relate the
low-energy constants (LECs) in the contact terms of
$\Sigma_c\bar{D}^{(*)}$ to those of nucleon systems, and reproduce
the $P_c(4312)$ and $P_c(4440)$ as loosely bound states. In the
second scenario, we vary the unknown LECs and find a small parameter
region in which  $P_c(4312)$, $P_c(4440)$ and $P_c(4457)$ can
coexist as molecular states. In the third scenario, we include the
coupled-channel effect on the basis of scenario II, and notice that
the three $P_c$ states can be reproduced as molecular states
simultaneously in a large region of parameters. Our analytical
results can be used for the chiral extrapolations in lattice QCD.
With the lattice QCD results in the future as inputs, the
identification of the $P_c$ states and predictions for other systems
would be more reliable.
\end{abstract}

\maketitle

\thispagestyle{empty}

\section{Introduction}\label{sec:intro}
The multiquark state has been a very important topic in hadron
physics for a long time (for recent reviews, see
Refs.~\cite{Chen:2016qju,Esposito:2016noz,Guo:2017jvc,Olsen:2017bmm,Liu:2019zoy}).
Since 2003, many ``XYZ'' states have been observed as the candidates
of tetraquark
states~\cite{Choi:2003ue,Aubert:2005rm,Ablikim:2013mio,Liu:2013dau}.
In 2015, the LHCb Collaboration discovered two pentaquark candidates
$P_c(4380)$ and $P_c(4450)$ in the $J/\psi p$ invariant mass
spectrum of $\Lambda_b\rightarrow J/\psi Kp$\cite{Aaij:2015tga}.
Very recently, the LHCb Collaboration reported the new results about
pentaquarks~\cite{Aaij:2019vzc}. The previously reported $P_c(4450)$
was resolved into two narrow states $P_c(4440)$ and $P_c(4457)$.
Besides, a new state $P_c(4312)$ was observed with 7.3$\sigma$
significance, and these three states are measured to be narrow. The
masses of these resonances lie below the thresholds of
$\Sigma_c\bar{D}$ and $\Sigma_c\bar{D}^{*}$, respectively. Thus,
they are good candidates of the molecular states. Before the
discovery of $P_c$ states, several groups predicted the existence of
the hidden charm molecular states with five
quarks~\cite{Wang:2011rga,Yang:2011wz,Wu:2012md}. Although there are
several alternative explanations, like tightly-bound
pentaquarks~\cite{Ali:2019npk,Maiani:2015vwa,Wang:2015ava,Weng:2019ynv}
and kinetic effect~\cite{Guo:2015umn}, the molecular state
explanation is more favorable. The recent QCD sum
rules~\cite{Chen:2019bip} and the one-boson-exchange (OBE)
model~\cite{Chen:2019asm,He:2019ify}, the local hidden gauge
approach~\cite{Xiao:2019aya} and the quark delocalization color
screening model~\cite{Huang:2019jlf} calculations also support the
molecule explanation for the $P_c(4312)$, $P_c(4440$) and
$P_c(4457)$ states. The productions and decays of the newly observed
pentaquark states were also investigated in
Refs.~\cite{Guo:2019fdo,Xiao:2019mvs,Cao:2019kst}.

The OBE model is widely used to study the nuclear
force~\cite{Yukawa:1935xg,Machleidt:1987hj}. For instance, the
deutron was well established as a hadronic molecular state in the
framework of OBE model~\cite{Machleidt:1987hj}. Additionally, in the
heavy flavor sector, the molecules with four quarks, five quarks and
six quarks were investigated with the OBE model as
well~\cite{Liu:2008tn,Li:2012bt,Chen:2015loa,Meng:2017fwb,Yang:2018amd}.

Chiral perturbation theory (ChPT) is used to build the modern the
nuclear force~\cite{Epelbaum:2008ga,Machleidt:2011zz}. The idea was
proposed by Weinberg~\cite{Weinberg:1990rz,Weinberg:1991um}. The
chiral expansion is performed to obtain the interaction kernel,
which is iterated to all orders by solving the Lippmann-Schwinger
equation or Schr\"{o}dinger equation. Compared with the OBE model,
the ChPT has the consistent power counting. The potential is
calculated order by order, and thus the error is estimable and
controllable. The same idea on nuclear force was also applied to the
heavy hadron systems. The interactions between two heavy mesons were
studied with the ChPT by taking the heavy quark symmetry (HQS) into
account~\cite{Valderrama:2012jv,Wang:2013kva,Baru:2015tfa,Liu:2012vd,Xu:2017tsr,Wang:2018atz}.
The $X(3872)$ and two $Z_b$ states were obtained as bound states
under this framework.

The HQS was also used to predict the partner states of $P_c(4312)$,
$P_c(4440$), $P_c(4457)$~\cite{Liu:2019tjn,Shimizu:2019ptd}. The HQS
is a good approximation when the heavy quark masses approach the
infinity. As we know, the heavy quark spin symmetry (HQSS) violation
effect will lead to mass splittings in the charmed sector between
the HQSS multiplets. The impacts of the HQS breaking effect on the
heavy molecular states are rarely estimated. The molecular states
are very shallow bound states and they are subtle to the behaviors
of the potentials.  Thus, the effect of HQSS violation on the
interactions between two heavy hadrons needs to be carefully
considered, especially in the charmed sector.

In this work, we derive the potentials between $\Sigma_{c}$ and
$\bar{D}^{(*)}$ in the framework of heavy hadron chiral perturbation
theory (HHCPT). We try to reproduce the the newly observed
$P_c(4312)$, $P_c(4440$) and $P_c(4457)$ as the molecular states. In
Sec.~\ref{sec:lag}, we discuss the Weinberg's formalism and
construct the Lagrangians. In Sec.~\ref{sec:pttl}, we perform the
Feynman diagrams calculation in the framework of the HHChPT and
obtain the analytical results of the effective potentials between
$\Sigma_{c}$ and $\bar{D}^{(*)}$ to the next-to-leading order. We
give some discussions about the $\Sigma_{c}\bar{D}^{(*)} $
potentials in the heavy quark limit in Sec.~\ref{sec:hqs}. In
Sec.~\ref{sec:numer}, we use three scenarios to give the numerical
results. A brief summary is given in Sec.~\ref{sec:sum}. In
Appendix~\ref{app:matrix}, we list the matrix elements of some
operators. In Appendix~\ref{app:integral}, we present the loop
integral functions we used.

\section{Weinberg's Formalism and Effective Lagrangians}\label{sec:lag}

In the framework of the HHChPT, the amplitudes are expanded in
powers of $\epsilon=q/\Lambda_{\chi}$, where $q$ is either the
momenta of Goldstone bosons or the residual momenta of the matter
fields, and $\Lambda_{\chi}$ is the chiral symmetry breaking scale.
For the singly heavy hadrons, the mass splitting $\delta$ between
the heavy quark multiplets is not vanishing in the chiral limit.
Thus, we adopt the small scale expansion in this
work~\cite{Hemmert:1997wz}, where the mass spitting $\delta$ is
regarded as another small scale. The amplitudes are also expanded in
powers of $\delta/m_{c}$. $m_{c}$ is the heavy quark mass, which is
treated as another large scale.

The expansion is organized according to the power counting given by
Weinberg~\cite{Weinberg:1990rz,Weinberg:1991um}. The order of a
diagram $\nu$ reads,
\begin{equation}
\nu=2L+2-\frac{E_{n}}{2}+\sum V_{i}\Delta_{i},\quad
\Delta_i=d_i+n_i/2-2, \label{pc}
\end{equation}
where $L$ and $E_n$ are the numbers of loops and external lines of
matter fields, respectively. For the $\Sigma_c \bar{D}^{(*)}$
potential, $E_n=2$. $V_i$ is the number of the vertices with order
$\Delta_i$. $d_i$ and $n_i$ are the numbers of the derivatives and
matter field lines, respectively.

In the Weinberg's formalism, only two particle irreducible (2PIR)
graphs are considered. The amplitudes of the box diagrams would be
enhanced by the pinch singularities, which would destroy the power
counting in Eq.~(\ref{pc}). The pinch singularity originates from
the two intermediate on-shell matter fields. Thus, one can recover
the power counting by excluding the two particle reducible (2PR)
contributions. The amplitudes we get in this way serve as the kernel
of Lippmann-Schwinger equation or Schr\"{o}dinger equation. The tree
level one-pion exchange diagrams would be iterated to generate the
2PR contributions automatically by solving the Lippmann-Schwinger
equation or Schr\"{o}dinger equation.

In order to remove the 2PR contributions in the box diagrams,
Ordonez \etal adopted the time-ordered perturbation
theory~\cite{Ordonez:1995rz}, while Kaiser \etal removed the
contributions from the poles of the intermediate matter fields when
performing the loop integrals~\cite{Kaiser:1997mw}. Here, we use the
principal integral method, which is equivalent to the scheme used in
Ref.~\cite{Kaiser:1997mw}. The details can be found in
Appendix~\ref{app:integral}.

In order to construct the chiral Lagrangians, we introduce the pion
fields,
\begin{eqnarray}
\phi =\sqrt{2}\left(\begin{array}{cc}
\frac{\pi^{0}}{\sqrt{2}} & \pi^{+}\\
\pi^{-} & -\frac{\pi^{0}}{\sqrt{2}}
\end{array}\right),~ U=u^{2}=\exp\left(i\frac{\phi(x)}{F_{0}}\right).
\end{eqnarray}
The chiral connection $\Gamma_{\mu}$ and the axial vector current
$u_{\mu}$ are defined as
\begin{eqnarray}
\Gamma_{\mu}=\frac{1}{2}[u^{\dagger},\partial_{\mu}u],\quad
u_{\mu}=\frac{i}{2}\{u^{\dagger},\partial_{\mu}u\}.
\end{eqnarray}
The multiplets of $\Sigma_c^{(*)}$ are denoted as
\begin{eqnarray}
\Sigma_{c}=\left(\begin{array}{cc}
\Sigma_{c}^{++} & \frac{\Sigma_{c}^{+}}{\sqrt{2}}\\
\frac{\Sigma_{c}^{+}}{\sqrt{2}} & \Sigma_{c}^{0}
\end{array}\right),~\Sigma_{c}^{*\mu}=\left(\begin{array}{cc}
\Sigma_{c}^{*++} & \frac{\Sigma_{c}^{*+}}{\sqrt{2}}\\
\frac{\Sigma_{c}^{*+}}{\sqrt{2}} & \Sigma_{c}^{*0}
\end{array}\right)^{\mu}.
\end{eqnarray}
Their chiral covariant derivative is
$D_{\mu}\Sigma^{(*)}_c=\partial_{\mu} \Sigma^{(*)}_c
+\Gamma_{\mu}\Sigma^{(*)}_c+\Sigma^{(*)}_{c}\Gamma_{\mu}^T$. The
leading order chiral Lagrangians for the $\Sigma_c^{(*)}$ read,
\begin{eqnarray}
{\cal L}_{\Sigma_{c}^{*}\phi}^{(0)}&=&\text{Tr}[\bar{\Sigma}_{c}(i\slashed{D}-M_{\Sigma_{c}})\Sigma_{c}]+ \text{Tr}[\bar{\Sigma}_{c}^{*\mu}[-g_{\mu\nu}(i\slashed{D}-M_{\Sigma_{c}^{*}})\nonumber \\
&+&i(\gamma_{\mu}D_{\nu}+\gamma_{\nu}D_{\mu})-\gamma_{\mu}(i\slashed{D}+M_{\Sigma_{c}^{*}})\gamma_{\nu}]\Sigma_{c}^{*\nu}]\nonumber \\
    &+&g_1\text{Tr}[\bar{\Sigma}_{c}\gamma^{\mu}\gamma_{5}u_{\mu}\Sigma_{c}]+g_3\text{Tr}[\bar{\Sigma}_{c}^{*\mu}u_{\mu}\Sigma_{c}]+\text{H.c}.\nonumber \\
    &+&g_5\text{Tr}[\bar{\Sigma}_{c}^{*\nu}\gamma^{\mu}\gamma_{5}u_{\mu}\Sigma_{c\nu}^{*}],
    \label{lag:siglo}
\end{eqnarray}
where $\text{Tr}[...]$ represents the trace in flavor space. Since
the $\Sigma_c$ and $\Sigma_c^{*}$ are the degenerate states in the
heavy quark limit, we can define the superfield as
\begin{eqnarray}
\psi^{\mu}  &=&\mathcal{B}^{*\mu}-\sqrt{\frac{1}{3}}(\gamma^{\mu}+v^{\mu})\gamma^{5}\mathcal{B},\nonumber \\
\bar{\text{\ensuremath{\psi}}}^{\mu}
&=&\bar{\mathcal{B}}^{*\mu}+\sqrt{\frac{1}{3}}\bar{\mathcal{B}}\gamma^{5}(\gamma^{\mu}+v^{\mu}),
\end{eqnarray}
where $\mathcal{B}^{*}$ stands for the $\Sigma_c^{(*)}$ fields after
heavy baryon reduction. The leading order Lagrangian in
Eq.~(\ref{lag:siglo}) can be rewritten as
\begin{eqnarray}\label{eq:SigmaLO}
{\cal L}_{\Sigma_{c}\phi}^{(0)}=&-&\text{Tr}[\bar{\psi}^{\mu}iv\cdot D\psi_{\mu}]+ig_{a}\epsilon_{\mu\nu\rho\sigma}\text{Tr}[\bar{\psi}^{\mu}u^{\rho}v^{\sigma}\psi^{\nu}]\nonumber \\
&+&
i\frac{\delta_{a}}{2}\text{Tr}[\bar{\psi}^{\mu}\sigma_{\mu\nu}\psi^{\nu}].
\label{lag:siglohq}
\end{eqnarray}
The third term in Eq.~(\ref{lag:siglohq}) accounts for the HQS
violation effect for the charmed baryons. $\delta_a=
M_{\Sigma_c^*}-M_{\Sigma_c}$ denotes the mass splitting. Comparing
Eq.~(\ref{lag:siglohq}) with Eq.~(\ref{lag:siglo}), one can easily
get
\begin{equation}
 g_{1}=-\frac{2}{3}g_{a},\quad g_{3}=-\sqrt{\frac{1}{3}}g_{a},\quad g_{5}=g_{a}.
\end{equation}

We introduce the superfield $\tilde{H}$ to denote the $\bar{D}$ and
$\bar{D}^*$ fields,
\begin{eqnarray}
&\tilde{H}  =(\tilde{P}_{\mu}^{*}\gamma^{\mu}+i\tilde{P}\gamma_{5})\frac{1-\slashed{v}}{2},\nonumber \\
&\bar{\tilde{H}}    =\frac{1-\slashed{v}}{2}(\tilde{P}_{\mu}^{*\dagger}\gamma^{\mu}+i\tilde{P}^{\dagger}\gamma_{5}),\\
& \tilde{P}=\left(\begin{array}{c}
\bar{D}^{0}\\
\bar{D}^{-}
\end{array}\right),\quad\tilde{P}^{*\mu}=\left(\begin{array}{c}
\bar{D}^{*0}\\
\bar{D}^{*-}
\end{array}\right),
\end{eqnarray}
where we use the ``$\tilde{X}$" to label the antiparticles. The
$\tilde{P}$ and $\tilde{P}^*_{\mu}$ are the reduced heavy meson
$\bar D$ and $\bar{D}^*$ fields, respectively, which can be related
to the relativistic $D^{(*)}$ fields $\Phi^{(*)}$ by the following
relations
\begin{eqnarray}
\sqrt{M}\Phi^{(*)}&=&e^{iMv\cdot x}\tilde{P}^{(*)\dagger}, \nonumber \\
\sqrt{M}\Phi^{(*)\dagger}&=&e^{-iMv\cdot x}\tilde{P}^{(*)},
\end{eqnarray}
where $M$ is the $D^{(*)}$ meson mass. Their chiral covariant
derivative is $D_{\mu}\tilde{P}^{(*)}=\partial_{\mu} \tilde{P}^{(*)}
+\Gamma_{\mu} \tilde{P}^{(*)}$. The leading order Lagrangians for
$\bar{D}$ and $\bar{D}^*$ are
\begin{equation}
{\cal L}^{(0)}_{\bar{D}\phi}=-i\langle\bar{\tilde{H}}v\cdot
D\tilde{H}\rangle+g_{b}\langle\bar{\tilde{H}}u_{\mu}\gamma^{\mu}\gamma_{5}\tilde{H}\rangle-\frac{\delta_{b}}{8}\langle\bar{\tilde{H}}\sigma^{\mu\nu}\tilde{H}\sigma_{\mu\nu}\rangle
,\label{lag:dlohq}
\end{equation}
where $\langle ...\rangle $ denotes the trace in spinor space. The
third term in Eq.~(\ref{lag:dlohq}) represents the HQS violating
effect for the charmed mesons. $\delta_b=M_{\bar{D}^*}-M_{\bar{D}}$
is the mass splitting between $\bar{D}^*$ and $\bar{D}$.

Apart from Eqs.~(\ref{lag:siglohq}) and (\ref{lag:dlohq}), the
leading order Lagrangians also contain the contact terms,
\begin{eqnarray}
{\cal L}^{(0)}_{\text{contact}} &=&D_{1}\langle\bar{\tilde{H}}\tilde{H}\rangle\text{Tr}(\bar{\psi}^{\mu}\psi_{\mu}) \nonumber \\
&+&iD_{2}\epsilon_{\sigma\mu\nu\rho}v^{\sigma}\langle\bar{\tilde{H}}\gamma^{\rho}\gamma_{5}\tilde{H}\rangle\text{Tr}(\bar{\psi}^{\mu}\psi^{\nu})\nonumber \\
    &+&\tilde{D_{1}}\langle\bar{\tilde{H}}\tau^{i}\tilde{H}\rangle\text{Tr}(\bar{\psi}^{\mu}\tau^{i}\psi_{\mu}) \nonumber \\
    &+&i\tilde{D_{2}}\epsilon_{\sigma\mu\nu\rho}v^{\sigma}\langle\bar{\tilde{H}}\gamma^{\rho}\gamma_{5}\tau^{i}\tilde{H}\rangle\text{Tr}(\bar{\psi}^{\mu}\tau^{i}\psi^{\nu}).
\end{eqnarray}

In this work, the values of the parameters~\cite{Tanabashi:2018oca}
are taken as
\begin{eqnarray}
& m_{\pi}=0.139\text{ GeV},~ F_{\pi}=0.092\text{ GeV},\nonumber\\
 &\delta_a=0.064\text{ GeV}, ~\delta_b=0.141\text{ GeV},~\lambda=1\text{ GeV},
\end{eqnarray}
where $\lambda$ is the large scale we used to perform the small
scale expansion. The $g_a$ and $g_b$ are determined by the partial
decay widths of $D^{*}$ and $\Sigma_c^{*}$
\cite{Ahmed:2001xc,Meguro:2011nr,Tanabashi:2018oca}, respectively.
\begin{equation}
g_a=-1.47,\quad g_b=-0.59.
\end{equation}
The signs of $g_a$ and $g_b$ are determined by the quark model. The
LECs in the contact terms will be estimated in Sec.~\ref{sec:numer}.

\section{effective potentials}\label{sec:pttl}

In order to obtain the effective potential, we calculate the
scattering amplitude $\cal{M}$ first. If we use the standard
normalization and Feynman rules in the relativistic quantum field
theory, the effective potential $\mathcal{V}(\bm{q})$ in the
momentum space reads,
\begin{equation}
\mathcal{V}(\bm{q})=-{\mathcal{M}\over \sqrt{2M_12M_22M_32M_4}},
\end{equation}
where $M_{1,\dots,4}$ are the masses of the scattering particles.
The potential in coordinate space can be obtained by making the
Fourier transformation to the $\mathcal{V}(\bm{q})$. In order to
regularize the divergence in the Fourier transformation, we
introduce the Gauss regulator, ${\cal
F}(\bm{q})=\exp(-\bm{q}^{2n}/\Lambda^{2n})$
\textcolor{red}~\cite{Ordonez:1995rz,Machleidt:2011zz,Ren:2016jna}.
The potential in the coordinate space reads,
\begin{equation}
V(r)=\frac{1}{(2\pi)^{3}}\int
d^{3}\bm{q}\,e^{i\bm{q}\cdot\bm{r}}{\cal V}(\bm{q}){\cal F}(\bm{q}).
\end{equation}
In this work, we set $n=2$ and vary the cutoff $\Lambda$ from $0.5$
GeV to $0.8$ GeV.

\subsection{$\Sigma_c\bar{D}$ system}
\begin{figure}[!htp]
\centering
\includegraphics[width=0.17\textwidth]{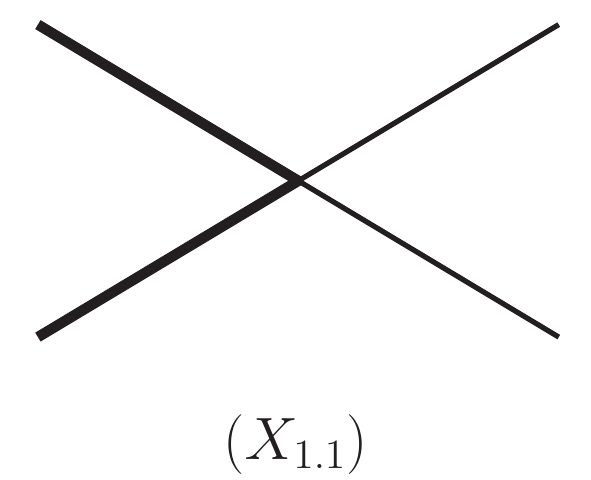}~~~~~~~
\caption{The leading order diagram for the $\Sigma_c\bar{D}$. At
this order, only the contact diagram $X_{1.1}$ contributes to the
effective potential. The thick solid and solid lines represent
$\Sigma_c$ and $\bar{D}$, respectively.}\label{fig:dlo}
\end{figure}

The leading order ($\mathcal{O}(\epsilon^0)$) potential of the
$\Sigma_c\bar{D}$ system comes from the contact diagram in
Fig.~\ref{fig:dlo}. There is no one-pion exchange diagram due to the
vanishing $\bar{D}\bar{D}\pi$ vertex. This vertex is forbidden by
the parity and angular momentum conservation. Thus, the leading
order potential of $\Sigma_c\bar{D}$ reads,
\begin{equation}
{\cal
V}_{\Sigma_{c}\bar{D}}^{X_{1.1}}=-D_{1}-\tilde{D}_{1}(2\bm{I}_1\cdot\bm{I}_2),
\end{equation}
where $\bm{I}_1\cdot\bm{I}_2$ is the isospin-isospin operator. The
superscript is the label of the Feynman diagram while the subscript
denotes the physical system.

\begin{figure*}[!htp]
\centering
\includegraphics[width=1\textwidth]{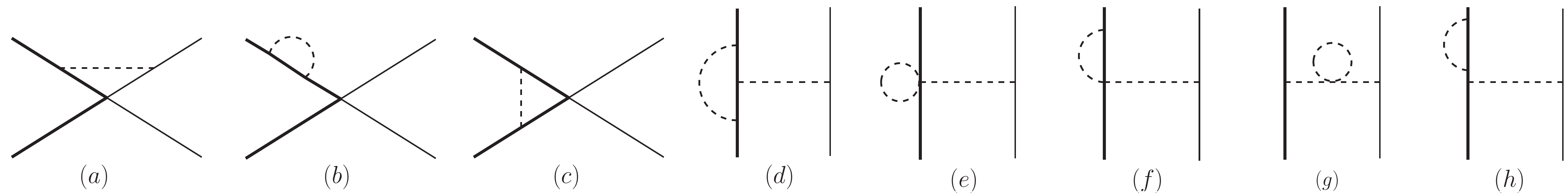}
\caption{The vertex correction and the wave function renormalization
diagrams at the next-to-leading order. Every graph represents a type
of Feynman diagrams with the same topological structure. Some
diagrams do not contribute due to the vanishing $\bar{D}\bar{D}\pi$
vertex.}\label{fig:remorm}
\end{figure*}

\begin{figure*}[!htp]
\centering
\includegraphics[width=1\textwidth]{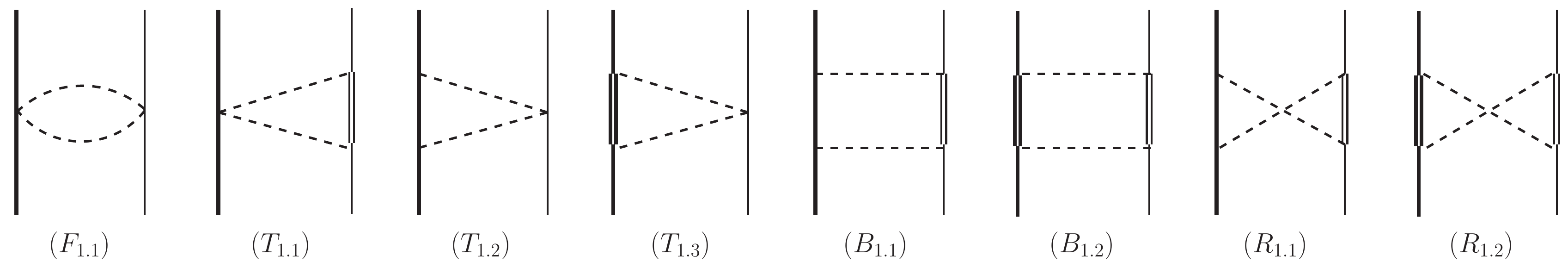}
\caption{The two pion exchange diagrams for the $\Sigma_c\bar{D}$.
There are one football diagram ($F_{1.1}$), three triangle diagrams
($T_{1.i}$), two box diagrams ($B_{1.i}$) and two crossed box
diagrams ($R_{1.i}$). The solid, thick solid, double solid, double
thick solid and dashed lines represent $\bar{D}$, $\Sigma_c$,
$\bar{D}^*$, $\Sigma_c^*$ and pion, respectively.}\label{fig:d2pion}
\end{figure*}

There are large amounts of contact terms contributing to the
effective potential at the next-to-leading order
($\mathcal{O}(\epsilon^2)$)~\cite{Wang:2018atz}. The LECs of these
vertices can be divided into the infinite part and finite part. The
infinite part can be used to absorb the divergence in the
$\mathcal{O}(\epsilon^2)$ loop diagrams. The renormalized vertices
(the finite part) will contribute to the effective potential. In our
calculations, we neglect the finite part of these vertices since we
have no experimental data as input to determine the LECs at
$\mathcal{O}(\epsilon^2)$.

There are loop diagrams contributing to the next-to-leading order
potential, which can be divided into two types. The first type is
the vertex correction and wave function renormalization diagrams in
Fig~\ref{fig:remorm}. Their contributions can be included when we
use the physical values of the parameters in Lagrangians. Another
type is the two-pion exchange diagrams in Fig~\ref{fig:d2pion},
including one football diagram, three triangle diagrams, two box
diagrams and two crossed box diagrams. The two-pion vertices in
these diagrams stem from the chiral connection terms in Eqs.
\eqref{eq:SigmaLO} and \eqref{lag:dlohq}. The one-pion vertices
arise from the axial coupling terms. In the calculation, we keep the
mass splittings from the HQS violation effect in the propagators.
The analytical results of these diagrams read
\begin{widetext}
\begin{eqnarray}
{\cal V}_{\Sigma_{c}\bar{D}}^{F_{1.1}}&=&\frac{J_{22}^{F}}{F_{\pi}^{4}}(\bm{I}_1\cdot\bm{I}_2),\\
{\cal V}_{\Sigma_{c}\bar{D}}^{T_{1.1}} &=&(\bm{I}_1\cdot\bm{I}_2)\frac{g_{b}^{2}}{F^{4}}[J_{34}^{T}(d-1)-(J_{33}^{T}+J_{24}^{T})\bm{q}^{2}](-\delta_{b}),\\
{\cal V}_{\Sigma_{c}\bar{D}}^{T_{1.2}} &=&(\bm{I}_1\cdot\bm{I}_2)\frac{-g_{1}^{2}}{4F_{\pi}^{4}}\left[(1-d)J_{34}^{T}+(J_{33}^{T}+J_{24}^{T})\bm{q}^{2}\right](0),\\
{\cal V}_{\Sigma_{c}\bar{D}}^{T_{1.3}}&=&(\bm{I}_1\cdot\bm{I}_2)\frac{g_{3}^{2}}{4F_{\pi}^{4}}\left[J_{34}^{T}(d-2)+(J_{33}^{T}+J_{24}^{T})\frac{2-d}{d-1}\bm{q^{2}}\right](-\delta_{a}),\\
{\cal V}_{\Sigma_{c}\bar{D}}^{B_{1.1}}  &=& (1-\bm{I}_1\cdot\bm{I}_2)\frac{-g_{1}^{2}g_{b}^{2}}{2F_{\pi}^{4}}\Bigg[J_{41}^{B}\frac{(d+1)(1-d)}{4}+(J_{42}^{B}+J_{31}^{B})\frac{d+1}{2}\bm{q^{2}}+J_{21}^{B}\frac{1}{4}\bm{q}^{2}\nonumber\\
&&~~~~~~~~~~~~~~~~~~~~~~~~~~~~~+(J_{43}^{B}+2J_{32}^{B}+J_{22}^{B})(-\frac{1}{4}\bm{q}^{4})\Bigg](0,-\delta_{b}),\\
{\cal V}_{\Sigma_{c}\bar{D}}^{B_{1.2}}&=&(1-\bm{I}_1\cdot\bm{I}_2)\frac{g_{3}^{2}g_{b}^{2}}{8F_{\pi}^{4}}\Bigg[J_{41}^{B}(d-2)(d+1)+(J_{42}^{B}+J_{31}^{B})\bm{q}^{2}\frac{2(d+1)(2-d)}{d-1}+J_{21}^{B}\bm{q^{2}}\frac{2-d}{d-1} \nonumber\\
&&~~~~~~~~~~~~~~~~~~~~~~+(J_{43}^{B}+2J_{32}^{B}+J_{22}^{B})\frac{d-2}{d-1}\bm{q^{4}}\Bigg](-\delta_{a},-\delta_{b}), \\
\mathcal{V}^{R_{i,j}}_{\Sigma_c\bar{D}}&=&
\mathcal{V}^{B_{i,j}}_{\Sigma_c\bar{D}}|_{J^B_x\rightarrow J^R_x,~
\bm{I}_1\cdot\bm{I}_2\rightarrow -\bm{I}_1\cdot\bm{I}_2},
\end{eqnarray}
\end{widetext}
where $J^T_{ij}$ are the  loop integrals defined in the
Appendix~\ref{app:integral}. They are the functions of $m_{\pi}$,
$\bm{q^2}$ and mass splittings $\delta_{a,b}$. We omit the
$m_{\pi}$, $\bm{q^2}$ for conciseness and give the specific mass
splittings at the end of every expression. The $\bm{q}$ is the
transferred three-momentum in the diagrams, and $d$ is the dimension
in the dimensional regularization. Since we focus on the $S$-wave
interactions, we take the replacement,
\begin{equation}
q^iq^j \rightarrowtail \delta^{ij}{1\over {d-1}}\bm{q^2}.
\end{equation}

\subsection{$\Sigma_c\bar{D}^*$ system}

\begin{figure}[!htp]
\centering
\includegraphics[width=0.35\textwidth]{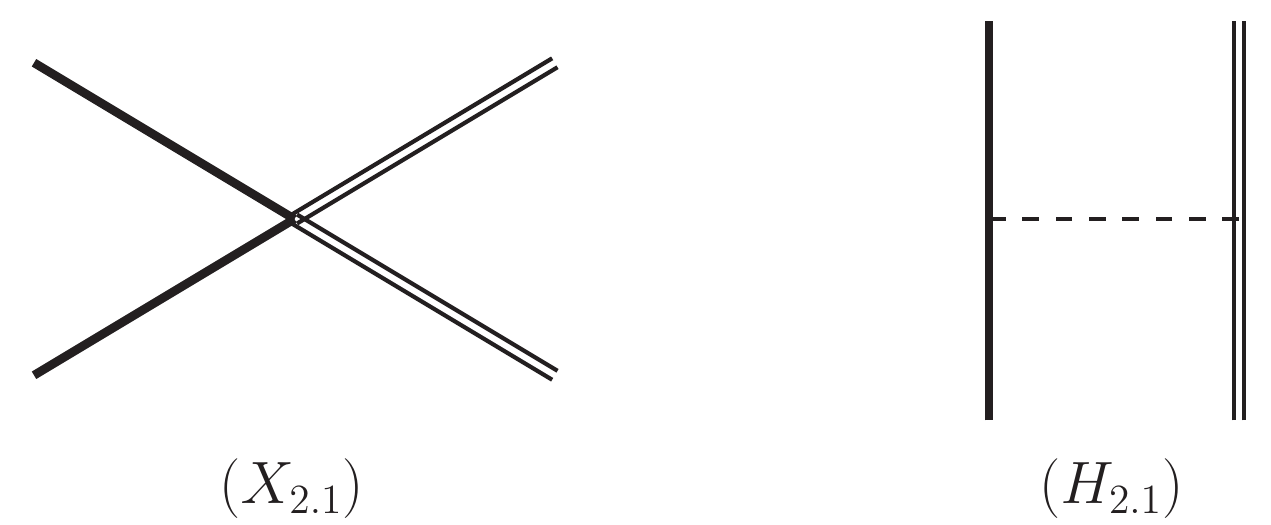}
\caption{The leading order diagrams for the $\Sigma_c\bar{D}^*$
system. At this order, the contact diagram $X_{2.1}$ and the
one-pion exchange diagram $H_{2.1}$ contributing to the effective
potential. Notations are the same as those in Fig.~\ref{fig:d2pion}.
}\label{fig:dslo}
\end{figure}

\begin{figure*}[!htp]
\centering
\includegraphics[width=1\textwidth]{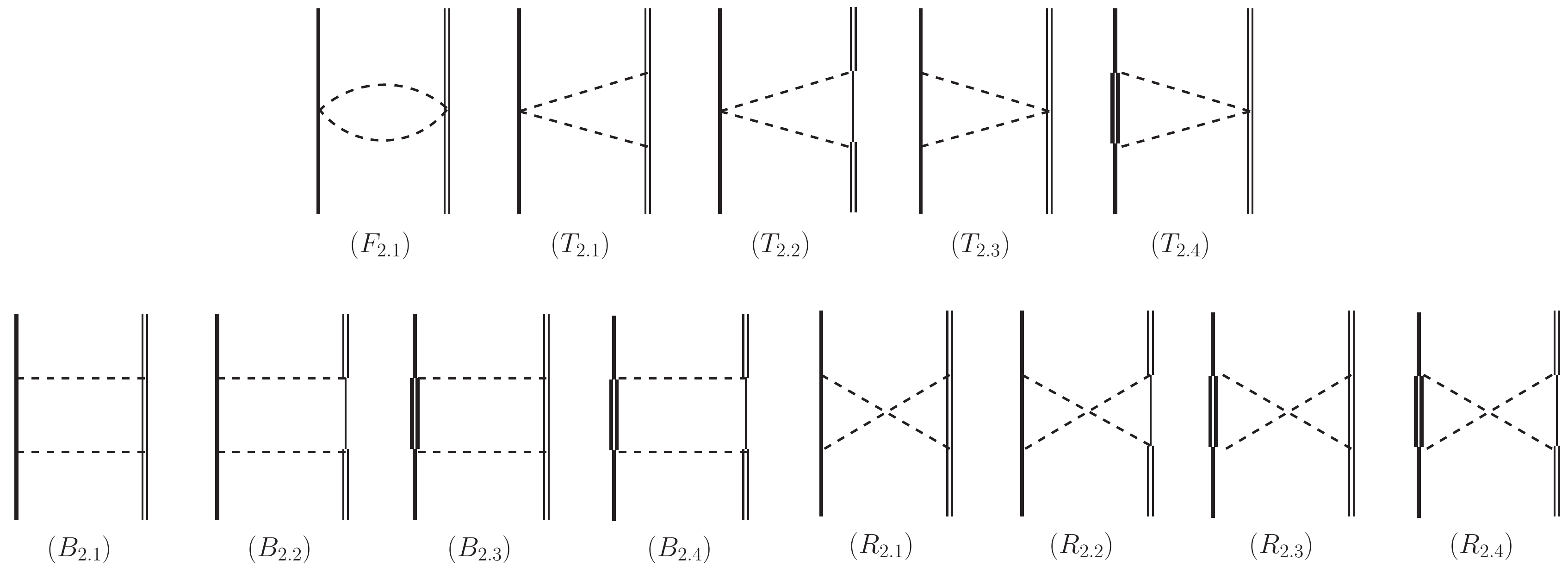}
\caption{The two-pion exchange diagrams for the $\Sigma_c\bar{D}^*$
system at the next-to-leading order. There is one football diagram
($F_{2.1}$), four triangle diagrams ($T_{2.i}$), four box diagrams
($B_{2.i}$) and four crossed box diagrams ($R_{2.i}$). Notations are
the same as those in Fig.~\ref{fig:d2pion}. }\label{fig:ds2pion}
\end{figure*}
For the $\Sigma_c\bar{D}^*$ system, the leading order potential is
generated from both the contact terms and the one-pion exchange
diagram. The analytical results read,
\begin{eqnarray}
{\cal V}_{\Sigma_{c}\bar{D}^*}^{X_{2.1}}&=&-\left(D_{1}+\frac{1}{3}D_{2}\bm{\sigma}\cdot\bm{T}\right)\nonumber \\
&~&-\left(\tilde{D}_{1}+\frac{1}{3}\tilde{D}_{2}\bm{\sigma}\cdot\bm{T}\right)(2\bm{I}_1\cdot\bm{I}_2),\\
{\cal V}_{\Sigma_{c}\bar{D}^{*}}^{H_{2.1}}
&=&-\frac{g_{b}g_{1}}{2F^{2}}\frac{(\bm{\sigma}\cdot\bm{q})(\bm{T}\cdot\bm{q})}{\bm{q}^{2}+m^{2}}(\bm{I_{1}}\cdot\bm{I_{2}}),
\end{eqnarray}
where $\bm{\sigma}$ is the Pauli matrix. The $\bm{T}\equiv
i\bm{\epsilon^*_4}\times\bm{\epsilon_2}$ is proportional to the spin
operator of $\bar{D}^*$. Thus, the $\bm{\sigma}\cdot\bm{T}$ terms
are the spin-spin interaction. For the $S$-wave potential, we can
make the following replacements in the effective potentials,
\begin{equation}
 \bm{\epsilon^*}_4 \cdot \bm{\epsilon}_2\rightarrowtail 1,\quad q^iq^j \rightarrowtail \delta^{ij}{1\over {d-1}}\bm{q^2}.
\end{equation}
Then, only the central terms and spin-spin terms survive in the
potential.

For the next-to-leading order potential, we neglect the finite part
from the $\mathcal{O}(\epsilon^2)$ contact terms again. The
renormalizations of the vertices, wave functions and masses are
included by using the physical values of  the coupling constants,
decay constants and masses just like we did in the previous
subsection.

At $\mathcal{O}(\epsilon^2)$, there are twelve two-pion exchange
diagrams which contribute to the effective potential in
Fig.~\ref{fig:ds2pion}. Their analytical results read,
\begin{widetext}
\begin{eqnarray}
{\cal V}_{\Sigma_{c}\bar{D}^{*}}^{F_{2.1}}&=&\frac{J_{22}^{F}}{F_{\pi}^{4}}(\bm{I}_1\cdot\bm{I}_2),\\
{\cal V}_{\Sigma_{c}\bar{D}^{*}}^{T_{2.1}}&=&(\bm{I}_1\cdot\bm{I}_2)\frac{g_{b}^{2}}{F_{\pi}^{4}}\left[2J_{34}^{T}-\frac{d-2}{d-1}\bm{q^{2}}(J_{33}^{T}+J_{24}^{T})\right](0),\\
{\cal
V}_{\Sigma_{c}\bar{D}^{*}}^{T_{2.2}}&=&(\bm{I}_1\cdot\bm{I}_2)\frac{-g_{b}^{2}}{F_{\pi}^{4}}\left[-J_{34}^{T}+(J_{33}^{T}+J_{24}^{T})\frac{1}{d-1}\bm{q^{2}}\right](\delta_{b}),
\\
{\cal V}_{\Sigma_{c}\bar{D}^{*}}^{T_{2.3}}&=&(\bm{I}_1\cdot\bm{I}_2)\frac{-g_{1}^{2}}{4F_{\pi}^{4}}\left[(1-d)J_{34}^{T}+(J_{33}^{T}+J_{24}^{T})\bm{q}^{2}\right](0),\\
{\cal V}_{\Sigma_{c}\bar{D}^{*}}^{T_{2.4}}&=&(\bm{I}_1\cdot\bm{I}_2)\frac{g_{3}^{2}}{4F_{\pi}^{4}}\left[J_{34}^{T}(d-2)+(J_{33}^{T}+J_{24}^{T})(\frac{2-d}{d-1})\bm{q^{2}}\right](-\delta_{a}),\\
{\cal V}_{\Sigma_{c}\bar{D}^{*}}^{B_{2.1}}&=&(1-\bm{I}_1\cdot\bm{I}_2)\frac{-g_{1}^{2}g_{b}^{2}}{8F^{4}}\Bigg[J_{41}^{B}(2(3-2d)+(J_{42}^{B}+J_{31}^{B})\bm{q}^{2}(4\frac{d-2}{d-1}+d)+J_{21}^{B}\frac{d-2}{d-1}\bm{q^{2}} \nonumber \\
&&+(J_{43}^{B}+2J_{32}^{B}+J_{22}^{B})(-\frac{d-2}{d-1}\bm{q}^{4})+J_{21}^{B}\frac{1}{d-1}\bm{q}^{2}\bm{T}\cdot\bm{\sigma}\Bigg](0,0),\\
{\cal V}_{\Sigma_{c}\bar{D}^{*}}^{B_{2.2}}&=&(1-\bm{I}_1\cdot\bm{I}_2)\frac{g_{1}^{2}g_{b}^{2}}{2F_{\pi}^{4}}\Bigg[J_{41}^{B}\frac{d+1}{4}+(J_{42}^{B}+J_{31}^{B})\bm{q}^{2}\frac{d+1}{2(1-d)}+J_{21}^{B}\frac{-1}{4(d-1)}\bm{q^{2}}\nonumber \\
&&+(J_{43}^{B}+2J_{32}^{B}+J_{22}^{B})\frac{1}{4(d-1)}\bm{q}^{4}+J_{21}^{B}\frac{-1}{4(d-1)}\bm{q^{2}}\bm{T}\cdot\bm{\sigma}\Bigg](0,\delta_{b}), \\
{\cal V}_{\Sigma_{c}\bar{D}^{*}}^{B_{2.3}} &=& (1-\bm{I}_1\cdot\bm{I}_2)\frac{g_{3}^{2}g_{b}^{2}}{32F_{\pi}^{4}}\Bigg\{ J_{41}^{B}(8\frac{d^{2}-2d+2}{d-1})+(J_{42}^{B}+J_{31}^{B})\bm{q}^{2}\left[-16(\frac{d-2}{d-1})^{2}-4\frac{d-2}{d-1}d\right]+J_{21}^{B}\left[-4(\frac{d-2}{d-1})^{2}\right]\bm{q}^{2}\nonumber\\
&&+(J_{43}^{B}+2J_{32}^{B}+J_{22}^{B})\frac{8(d-2)}{(d-1)^{2}}\bm{q}^{4}+J_{21}^{B}\frac{4}{(d-1)^{2}}\bm{q}^{2}\bm{T}\cdot\bm{\sigma}\Bigg\}(-\delta_{a},0),\\
{\cal V}_{\Sigma_{c}\bar{D}^{*}}^{B_{2.4}}&=&(1-\bm{I}_1\cdot\bm{I}_2)\frac{-g_{3}^{2}g_{b}^{2}}{8F_{\pi}^{4}}\Bigg[J_{41}^{B}(\frac{2}{d-1}-d)+(J_{42}^{B}+J_{31}^{B})\frac{2(d-2)(d+1)}{(d-1)^{2}}\bm{q}^{2}+J_{21}^{B}\frac{(d-2)}{(d-1)^{2}}\bm{q}^{2}\nonumber\\
&&+(J_{43}^{B}+2J_{32}^{B}+J_{22}^{B})\frac{2-d}{(d-1)^{2}}\bm{q}^{4}-J_{21}^{B}\frac{1}{(d-1)^{2}}\bm{q^{2}}\bm{T}\cdot\bm{\sigma}\Bigg](-\delta_{a},\delta_{b}),\\
\mathcal{V}^{R_{i,j}}_{\Sigma_c\bar{D}^*}&=&
\mathcal{V}^{B_{i,j}}_{\Sigma_c\bar{D}^*}|_{J^B_x\rightarrow J^R_x,~
\bm{I}_1\cdot\bm{I}_2\rightarrow -\bm{I}_1\cdot\bm{I}_2,~
\bm{\sigma}\cdot\bm{T}\rightarrow -\bm{\sigma}\cdot\bm{T}},
\end{eqnarray}
where the notations are the same as those for the $\Sigma_c\bar{D}$
system.
\end{widetext}

\section{The heavy quark symmetry}\label{sec:hqs}
In Secs.~\ref{sec:lag} and \ref{sec:pttl}, the LECs can be related
to one another by adopting the HQS. The HQS violation effect is
introduced through the mass splittings. If we ignore these mass
splittings in the loop diagrams, the HQS manifests itself.

The manifestation of the HQS could be clearer at the quark level. In
the heavy quark limit, the potential between the $\Sigma_c$ and
$\bar{D}^{(*)}$ arises from the interactions of their light degrees
of freedom. The heavy degrees of freedom are spectators. Their
interactions are suppressed by the heavy quark mass. The $S$-wave
interactions between the light diquark in the $\Sigma_c$ and the
light quark in the $\bar{D}^{(*)}$ can be expressed as,
\begin{equation}
V_{\mathrm{quark-level}}^{\mathrm{HQS}}=V_{a}+\tilde{V_{a}}\bm{l}_{1}\cdot
\bm{l}_{2},\label{eq:hqs-qm}
\end{equation}
where $\bm{l}_1$ and $\bm{l}_2$ are the spin operators of their
light degrees of freedom. Since we concentrate on the $S$-wave
interactions, only the spin-spin interaction and the central
potential exist. We can parameterize the potential at the hadron
level as
\begin{eqnarray}
V_{\Sigma_{c}\bar{D}}   &=&V_{1},\nonumber \\
V_{\Sigma_{c}\bar{D}^{*}}   &=&V_{2}+\tilde{V}_{2}\bm{S}_{1}\cdot \bm{S}_{2},\nonumber \\
V_{\Sigma_{c}^{*}\bar{D}}   &=&V_{3},\nonumber \\
V_{\Sigma_{c}^{*}\bar{D}^{*}}  &=&
V_{4}+\tilde{V}_{4}\bm{S}_{1}\cdot \bm{S}_{2}.
\end{eqnarray}
With the quark level interaction in the HQS, we can relate the
hadron level potentials with each other as follows,
\begin{eqnarray}
&&V_{1}=V_{2}=V_{3}=V_{4}=V_{a},\nonumber \\
&&\tilde{V}_{2}=\frac{2}{3}\tilde{V}_{a},\quad
\tilde{V}_{4}=\frac{1}{3}\tilde{V}_{a}.\label{eq:hq}
\end{eqnarray}
We give the matrix elements of $\bm{l}_1\cdot \bm{l}_2$ and
$\bm{S}_1\cdot \bm{S}_2$ in Table~\ref{tab:matrixele} and the
calculation details in Appendix~\ref{app:matrix}. Our analytical
results indeed satisfy the above expressions when $d\rightarrow 4$
and $\delta_{a,b}\rightarrow 0$.

With Eq.~(\ref{eq:hqs-qm}), we can get the potentials of the
$\Sigma_c^* \bar{D}$, $\Sigma_c^* \bar{D}^*$ and even the inelastic
channels in the heavy quark limit without calculating the loop
diagrams. We first extract the $V_a$ and $\tilde{V}_a$ from the
$\Sigma_{c}\bar{D}^{*}$ potentials as in Eq.~(\ref{eq:hq}). Then, we
calculate the matrix elements of $\bm{l}_1\cdot \bm{l}_2$ for the
corresponding channels. The potential can then be derived from the
quark-level interaction in Eq.~(\ref{eq:hqs-qm}). In our framework,
the leading order potentials satisfy the HQS, i.e., the leading
order potentials we derived from above procedures are equal to those
from Feynman diagrams.
\begin{table*}
    \caption{Matrix elements of $\bm{l}_1\cdot \bm{l}_2$ and $\bm{S}_1\cdot \bm{S}_2$ operators in the $\Sigma_c^{(*)}\bar{D}^{(*)}$ wave functions. $J$ denotes the total spins of the two hadrons.}\label{tab:matrixele}
\setlength{\tabcolsep}{5.2mm}
    \begin{tabular}{c|ccccccc}
        \toprule[1pt]\toprule[1pt]
        $\bar{\Sigma}_{c}^{(*)}\bar{D}^{(*)};J$ & $\Sigma_{c}\bar{D};\frac{1}{2}$ & $\Sigma_{c}\bar{D}^{*};\frac{1}{2}$ & $\Sigma_{c}\bar{D}^{*};\frac{3}{2}$ & $\bar{\Sigma}_{c}^{*}\bar{D};\frac{3}{2}$ & $\bar{\Sigma}_{c}^{*}\bar{D}^{*};\frac{1}{2}$ & $\bar{\Sigma}_{c}^{*}\bar{D}^{*};\frac{3}{2}$ & $\bar{\Sigma}_{c}^{*}\bar{D}^{*};\frac{5}{2}$\tabularnewline
        \midrule[1pt]
        $\langle\bm{l}_1\cdot\bm{l}_2\rangle$ & 0 & $-\frac{2}{3}$ & $\frac{1}{3}$ & 0 & $-\frac{5}{6}$ & $-\frac{1}{3}$ & $\frac{1}{2}$\tabularnewline
        $\langle\bm{S}_1\cdot\bm{S}_2\rangle$ & 0 & $-1$ & $\frac{1}{2}$ & 0 & $-\frac{5}{2}$ & $-1$ & $\frac{3}{2}$\tabularnewline
        \bottomrule[1pt]\bottomrule[1pt]
    \end{tabular}
\end{table*}

In the heavy quark limit, the analytical results of the box diagrams
become more concise. All the mass splittings between heavy quark
multiplets vanish with the HQS, and all the box diagrams with the
pinch singularities become the 2PR diagrams. With the expression of
$J^B_x$ in Appendix~\ref{app:integral}, there exists the relation
$J^B_x=-J^R_x$ when $\delta_a=\delta_b=0$. The total potentials of
the box diagrams and crossed box diagrams are
\begin{equation}
\mathcal{V}^R+\mathcal{V}^B=\mathcal{V}_C(\bm{I}_1\cdot\bm{I}_2)+\mathcal{V}_S(\bm{\sigma}\cdot\bm{T}).
\end{equation}
The results are similar to those for the nuclear force
\textcolor{red}~\cite{Machleidt:2011zz}.

\begin{figure*}[htp]
\centering
\begin{tabular}{ccc}
\includegraphics[width=0.32\textwidth]{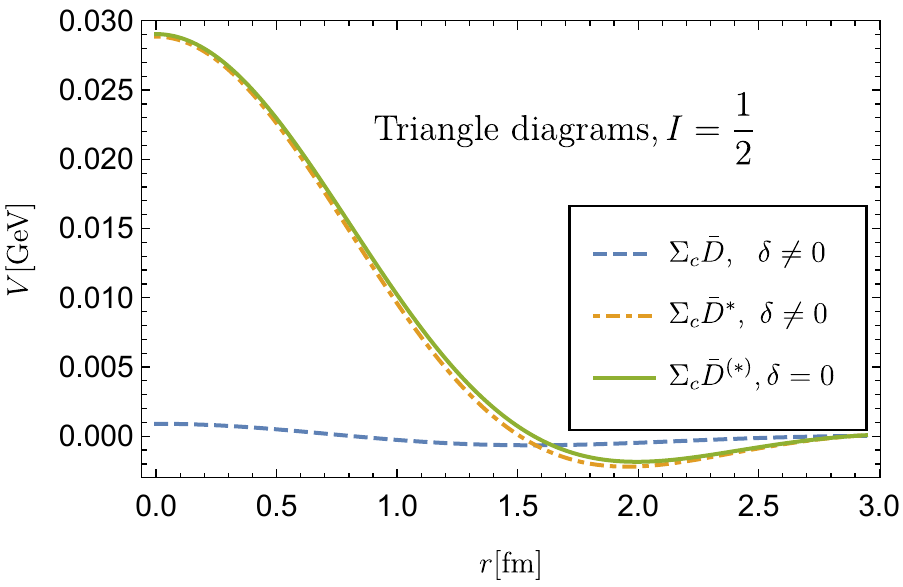}~&~
\includegraphics[width=0.32\textwidth]{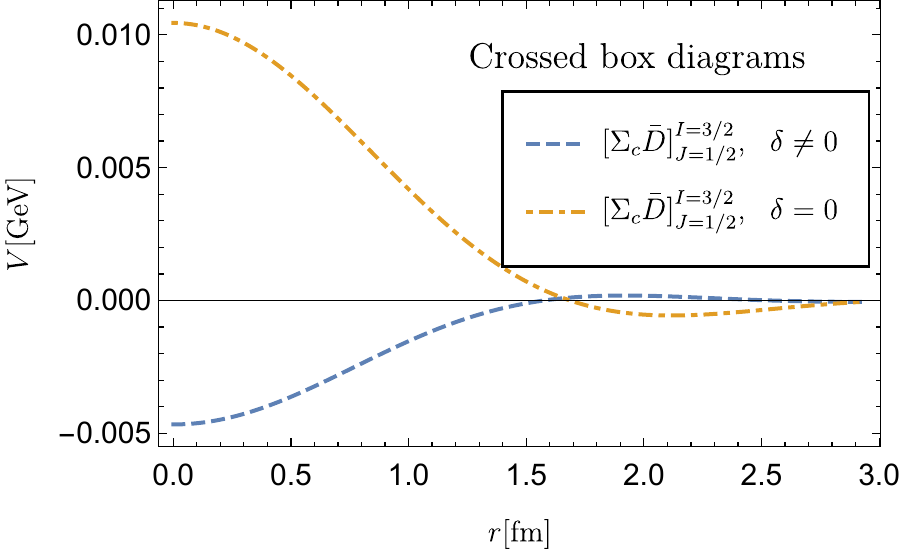}~&~
\includegraphics[width=0.32\textwidth]{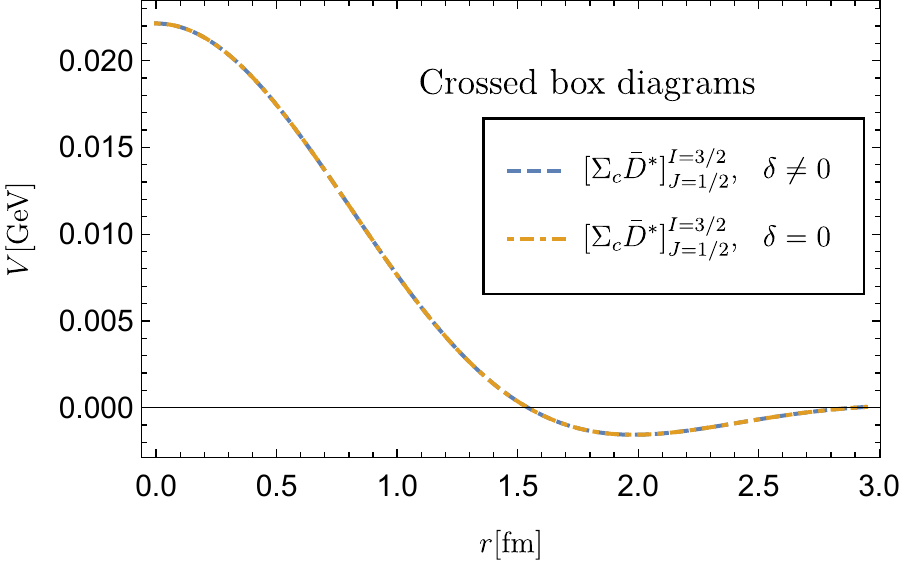}\\
\end{tabular}
\caption{The potentials of triangle diagrams and crossed box
diagrams. The $\delta=0$ in the legends denotes that we ignore the
mass splittings, which corresponds to the result in the heavy quark
limit. The $\delta\neq 0$ denotes that we keep the mass splittings
from the HQS violation effect.}\label{fig:checkhq}
\end{figure*}

However, the HQS is still an approximation when the heavy quark mass
is not infinite. Whether the HQS is good enough in calculating the
heavy hadron potential to obtain the bound states needs to be
considered carefully. In our calculation, we keep the mass
differences stemming from the HQS breaking effect. For the triangle
diagrams, there is no spin-spin interaction. Thus in the heavy quark
limit, the $\Sigma_c\bar{D}$ and $\Sigma_c\bar{D}^*$ should have the
same potential. We give the potentials of the triangle diagrams for
the $I={1\over 2}$ systems in Fig.~\ref{fig:checkhq}, from which we
see that the $\Sigma_c\bar{D}^*$ potential in the heavy quark limit
is very close to its real potential. However, the $\Sigma_c\bar{D}$
potential in the heavy quark limit deviates a lot from its real
potential. Similar results are obtained for the crossed box diagrams
in Fig.~\ref{fig:checkhq} as well. For the $\Sigma_c\bar{D}^*$
system, the potential in the HQS is a good approximation of its real
potential. However, the real $\Sigma_c\bar{D}$ attractive potential
becomes repulsive when the HQS is adopted. In the triangle or
crossed box diagrams of the $\Sigma_{c}\bar{D}$ system, ignoring the
HQS violation effect will make the potential in the coordinate space
change by about 0.02-0.03 GeV. Our numerical results in
Sec.~\ref{sec:numer} indicate that the minimum of the potential
function that generates the loosely bound state is from -0.06 GeV to
-0.15 GeV. The correction from the HQS violation is not negligible.
Therefore, it may be misleading to adopt the HQS in calculating the
charmed hadron potentials, at least for the $\Sigma_c\bar{D}$
system.

We notice that the HQS violation effect is more significant for the
$\Sigma_c\bar{D}$ system than that in the $\Sigma_c\bar{D}^*$ case.
We take the triangle diagrams as an example to illustrate the
reason. In Fig.~\ref{fig:d2pion}, the graphs $(T_{1.1})$ and
$(T_{1.3})$ are two diagrams with the HQS violation effect for the
$\Sigma_c\bar{D}$ system. Their intermediate states are $\Sigma_c^*$
and $\bar{D}^*$, respectively, which are both heavier than the
corresponding external particles. The HQS violation effect will
deviate the potential in the same direction. Thus, the HQS violation
effect from different diagrams are constructive. For the
$\Sigma_c\bar{D}^*$ system, the intermediate state can be either
heavier or lighter than its corresponding external field. The HQS
breaking (HQSB) effect would cancel with each other. Thus, we can
infer that the HQS violation effect from mass splittings is also
significant for the $\Sigma_c^*\bar{D}^*$ system.

The HQS violation effect can also be investigated at the quark
level. To this end, we present this effect as follows,
\begin{equation}
V^{\mathrm{HQSB}}_{\mathrm{quark-level}}={V_c\over
m_c}\bm{l}_1\cdot\bm{ h}_2+{V_d\over m_c}\bm{l}_2\cdot\bm{
h}_1+{V_e\over m_c^2}\bm{h}_1\cdot\bm{ h}_2,\label{eq:HQSB}
\end{equation}
where $V_c$, $V_d$ and $V_e$ are the functions used to parameterize
the potential. The first and second terms are the interaction
between the light and heavy degrees of freedom, which are suppressed
by the $1/m_c$. The third term is the interaction between the heavy
degrees of freedom, which is the higher order contribution in the
heavy quark expansion. For the $\Sigma_{c}\bar{D}$ system, we
calculate the matrix elements of three spin-spin operators in
Eq.~(\ref{eq:HQSB}),
\begin{eqnarray}
\langle \bm{l_1}\cdot\bm{ h}_2 \rangle =\langle \bm{l}_2\cdot\bm{
h}_1 \rangle=\langle \bm{h}_1\cdot\bm{ h}_2 \rangle =0.
\end{eqnarray}
The HQS violation effect vanishes, which seems to be contradictory
with the conclusion from directly calculating Feynman diagrams. The
HQS violation effect in calculating the loop diagrams arises from
the mass splittings in the propagators. It is hard to include this
effect in the quark model. Quark model can only give the analytical
terms which are the polynomials of $m_{\pi}^2$ or $\delta$. However,
the loop diagrams in ChPT can generate the nonanalytical structures
such as the logarithmic terms. The quark level HQS violation effect
in Eq.~(\ref{eq:HQSB}) is more likely to appear in the LECs at the
hadron level.

\begin{figure}[htp]
\centering
\includegraphics[width=0.48\textwidth]{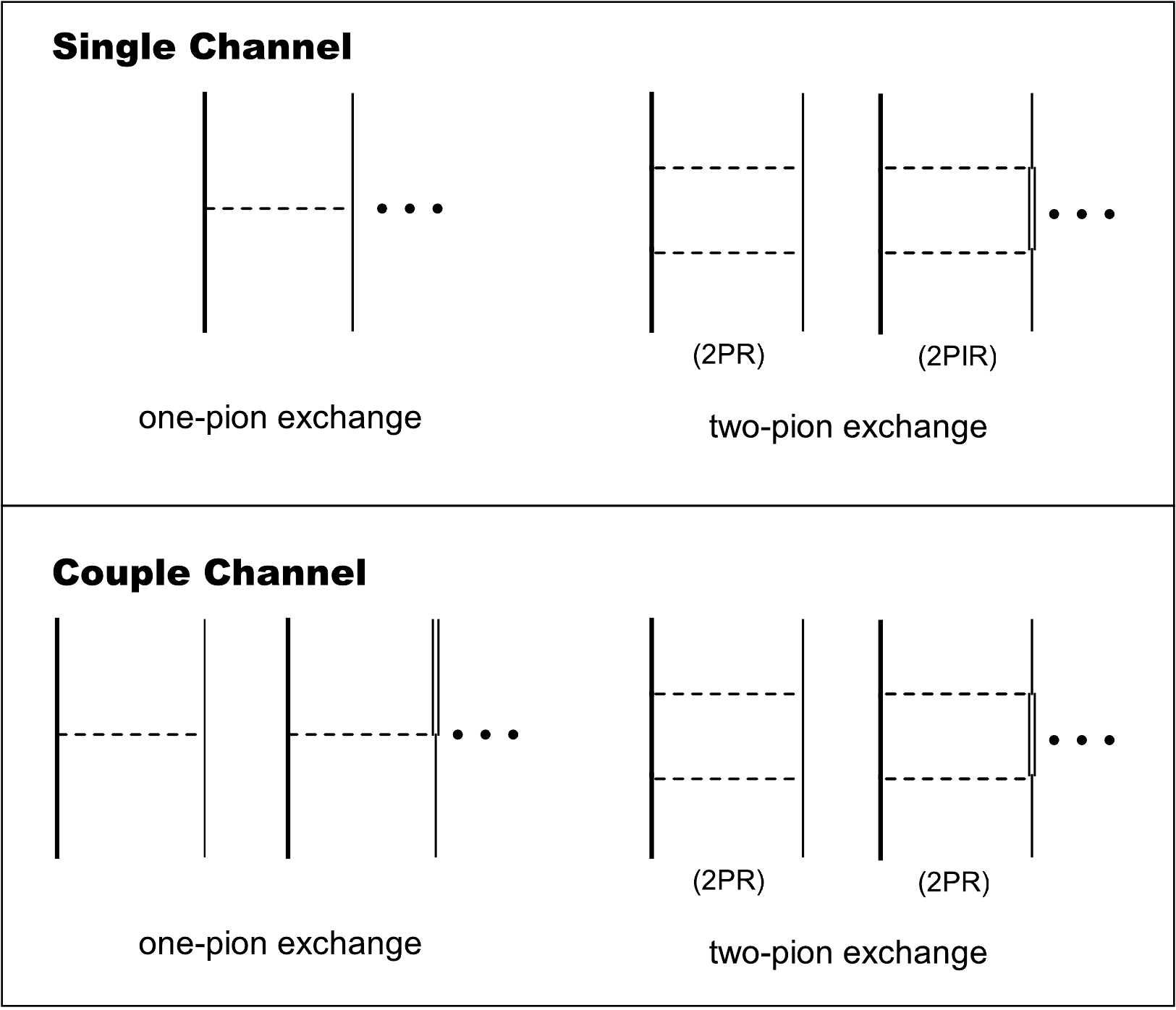}
\caption{The different treatments to the boxed diagram in
single-channel and coupled-channel calculations. }\label{fig:cc}
\end{figure}

The heavy quark limit for the box diagrams is more tricky. When the
intermediate states of the box diagrams are the HQS partner states
of the external fields, we calculate these diagram directly. They
have no pinch singularities due to the existence of the mass
splittings. If we decrease the mass splittings to zero, these
amplitudes will blow up and these diagrams become two particle
reducible. One way to eliminate these singularities is to remove the
2PR contributions. Meanwhile, we shall solve the coupled-channel
Schr\"{o}dinger equation as illustrated in Fig.~\ref{fig:cc}. The
one-pion exchange inelastic scattering diagrams at the tree level
will be iterated to generate the 2PR contributions automatically.
Thus, it is illegitimate to take the mass splittings in our
analytical results of the box diagrams to zero directly.

\section{Numerical results}\label{sec:numer}
In our analytical results, there are four unknown LECs from the
contact terms, $D_1$, $D_2$, $\tilde{D}_1$ and $\tilde{D}_2$. These
LECs should be determined by fitting the experimental data. However,
there does not exist any $\Sigma_c$ and $\bar{D}^{(*)}$ scattering
data now. Therefore, we choose three scenarios to present the
numerical results in this section. In scenario I, we determine these
LECs from the nucleon scattering with the help of quark model. In
scenario II, we will assume the three structures observed in LHCb
corresponding to three molecular states and check whether they can
coexist in our framework. In scenario III, we will include the
coupled-channel effect on the basis of scenario II.


\subsection{Scenario I}

There are two motivations to introduce the contact terms. First,
some heavy mesons exchanged between $\Sigma_c$ and $\bar{D}^{(*)}$
like $\rho$ and $\omega$, are integrated out, and their
contributions are included in the contact terms. Second, the contact
terms will absorb the divergence in the loops and remove the scale
dependence. The contact terms will cancel the infinity in the loop
diagrams. Thus, the values of renormalized contact terms depend on
chiral truncation order. Meanwhile, the contact terms will make the
potential scale independent. The values of the contact terms will
depend on the regularization schemes, types of regulator and values
of cutoff. The specific values of the contact terms for the nucleon
system in literatures also vary due to above
reasons~\cite{Ordonez:1995rz,Epelbaum:2004fk,Machleidt:2011zz}.

For the nucleon system, the leading order contact Lagrangian and the
potential are written as
\begin{eqnarray}
&{\cal L}_{NN}^{(0)}=-\frac{1}{2}C_{S}\bar{N}N\bar{N}N-\frac{1}{2}C_{T}\bar{N}\bm{\sigma}N\cdot\bar{N}\bm{\sigma}N, \\
&{\cal V}_{NN}=C_{S}+C_{T}\bm{\sigma}\cdot\bm{\sigma}.
\end{eqnarray}
There are two independent LECs, $C_S$ and $C_T$. The isospin-isospin
interaction is absorbed through Fierz rearrangement. Since we use
the dimensional regularization in calculations, we choose the values
determined in Ref.~\cite{Machleidt:2011zz}, in which the same
regularization scheme was used. We take the values
\begin{equation}
C_S=-99.43\text{ GeV}^{-2},\quad C_T=6.95\text{ GeV}^{-2}.
\end{equation}

In order to relate the contact terms in the nucleon system to those
of the $\Sigma_c\bar{D}^{(*)}$ systems, we make use of the contact
interaction at the quark level,
\begin{eqnarray}
{\cal
L}_{\mathrm{quark}}=-\frac{1}{2}c_{s}\bar{q}q\bar{q}q-\frac{1}{2}c_{t}(\bar{q}\bm{\sigma}q)\cdot(\bar{q}\bm{\sigma}q),
\end{eqnarray}
where $q=(u,d)^T$. $c_s$ and $c_t$ are the coupling constants at the
quark level. The interaction may arise from the heavy meson exchange
at the quark level. With the quark model, we can get the values of
the LECs for the $\Sigma_c\bar{D}^{(*)}$ systems,
\begin{eqnarray}
&D_{1}=-\frac{2}{9}C_{S}=22.1\text{ GeV}^{-2},\quad D_{2}=4C_{T}=27.8\text{ GeV}^{-2},\nonumber \\
& \tilde{D}_1=\tilde{D}_2=0.\label{eq:contactvalue1}
\end{eqnarray}

\begin{figure*}[htp]
\centering
\begin{tabular}{ccc}
\includegraphics[width=0.4\textwidth]{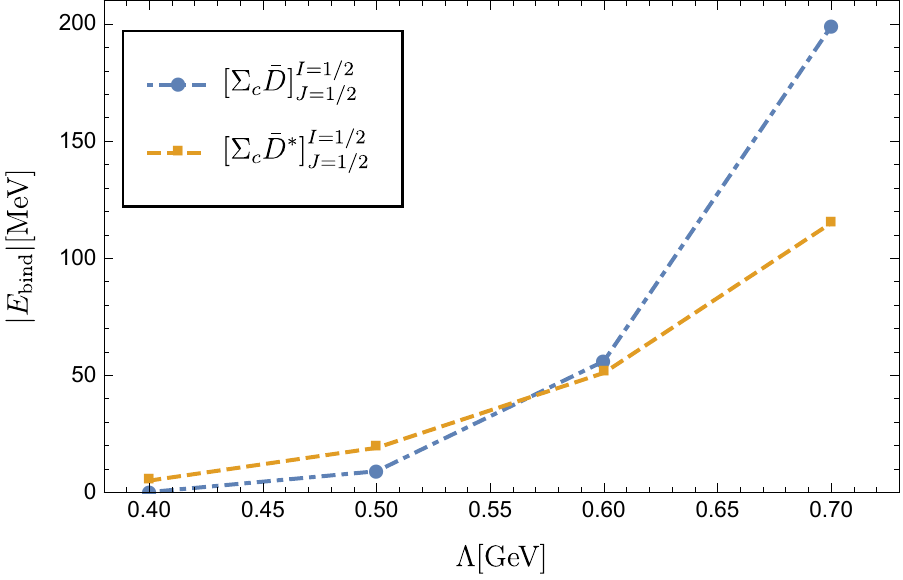}~&~
\includegraphics[width=0.4\textwidth]{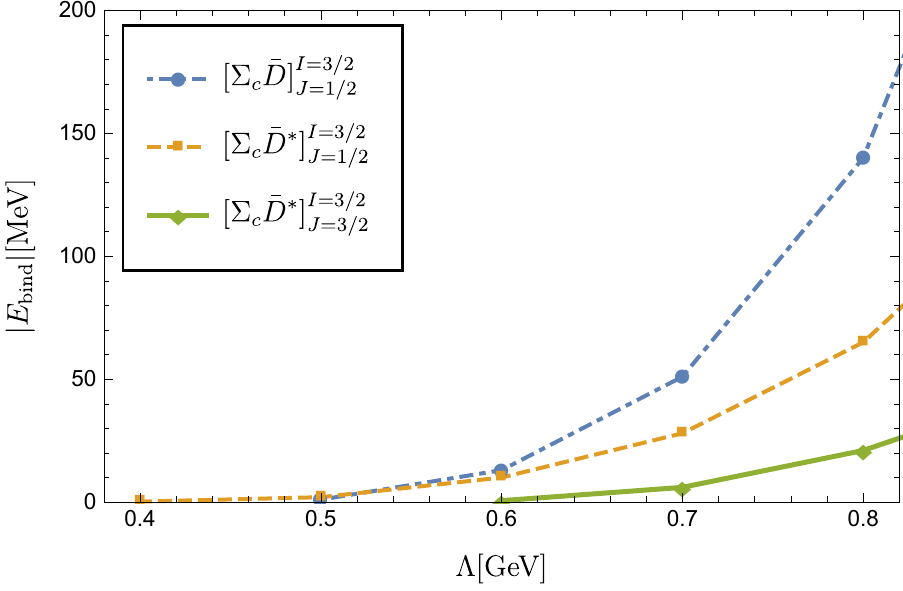}\\
\end{tabular}
\caption{The binding energies of the $\Sigma_c\bar{D}^{(*)}$ system.
The left one and the right one are the $I={1\over 2}$ systems and
$I={3\over 2}$ systems, respectively. The systems without binding
solutions are not given.}\label{fig:sov}
\end{figure*}

\begin{figure*}[htp]
\centering
\begin{tabular}{ccc}
\includegraphics[width=0.32\textwidth]{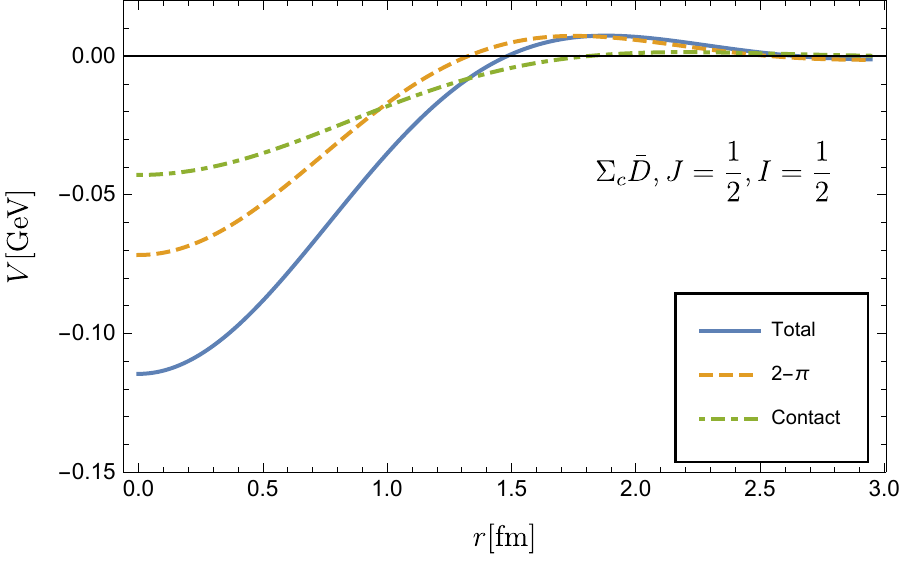}~&~
\includegraphics[width=0.32\textwidth]{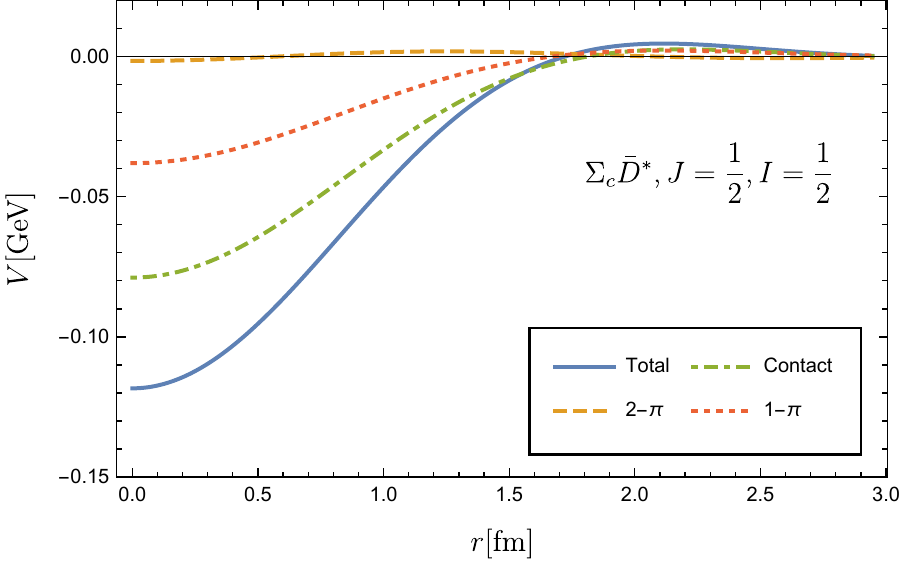}~&~
\includegraphics[width=0.32\textwidth]{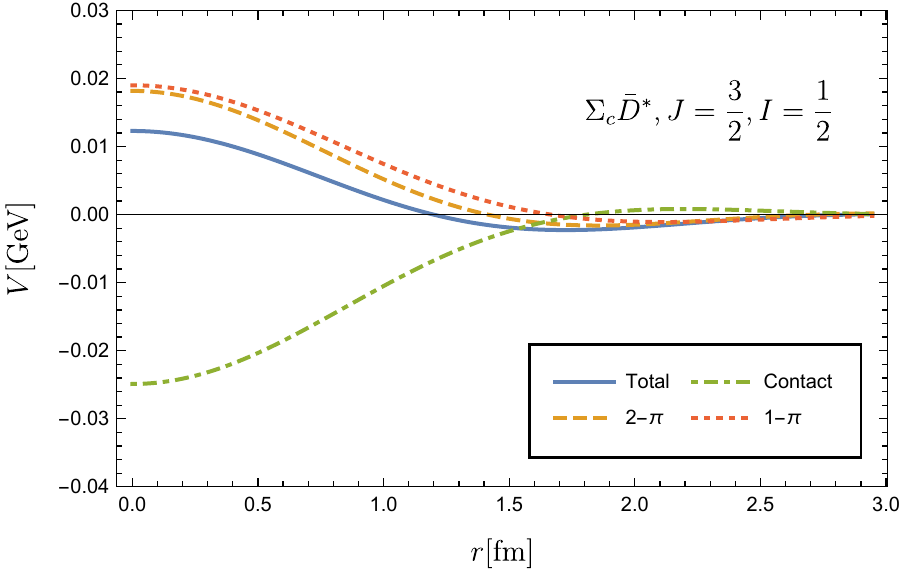}\\
\end{tabular}
\caption{The potentials for the $[\Sigma_c\bar{D}^{(*)}]^{I=1/2}$
systems in scenario I. The cutoff parameter $\Lambda=0.5$ GeV. The
LECs of the contact terms are taken from
Eq.~\eqref{eq:contactvalue1}. }\label{fig:ptl}
\end{figure*}

With these LECs, we can solve the Schr\"{o}dinger equation to obtain
some bound states. We vary the cutoff $\Lambda$ from $0.4$ GeV to
$0.8$ GeV. The binding solutions are given in Fig.~\ref{fig:sov}.
For the $I={1\over 2}$ system, we got binding solutions for the
$[\Sigma_c\bar{D}]_{J=1/2}$ and  $[\Sigma_c\bar{D}^*]_{J=1/2}$
systems. We reproduce the masses of $P_c(4312)$ and $P_c(4440)$ when
the cutoff is $0.5$ GeV. We present the potentials in
Fig.~\ref{fig:ptl}. The binding energy and the root mean square
radius (RMS) for the $[\Sigma_c\bar{D}]_{J=1/2}$ are $-9.21$ MeV and
$1.36$ fm, respectively. For the $[\Sigma_c\bar{D}^*]_{J=1/2}$, the
binding energy and the RMS are $-18.93$ GeV and $1.16$ fm,
respectively.

However, in this scenario, we can not reproduce the $P_c(4457)$. In
fact, the potential of $[\Sigma_c\bar{D}^*]_{J=3/2}^{I=1/2}$ is
repulsive as shown in Fig.~\ref{fig:ptl}. The scheme we used to
determine the LECs is rather rough. Therefore, we can not rule out
the possibility of the $P_c(4457)$ as a molecular state because of
the uncertainty of the LECs. For all three $I={3\over 2}$ systems,
there exist loosely bound states when we vary the cutoff $\Lambda$
from $0.4$ GeV to $0.8$ GeV. We give the results in
Fig.~\ref{fig:sov}.

\subsection{Scenario II}
One can use other phenomenological methods to evaluate the LECs,
such as heavy meson exchange model. But they also bring large
uncertainties. Meanwhile, we drop out the finite contributions from
$\mathcal{O}(\epsilon ^2)$ contact terms, which may also influence
our final results. In scenario II, we will adopt the general form of
contact terms and vary the LECs to search for the bound solutions.

In this scenario, we will focus on the $I=1/2$ systems since the
three $P_c$ states were all observed in the $J/\psi p$ invariant
mass spectrum. Thus, there are only two independent contact terms.
We parameterize the contact interaction of
$[\Sigma_c\bar{D}^{(*)}]^{I=1/2}$ as,
\begin{equation}
{\cal V}_{\Sigma_{c}\bar{D}}^{X_{1.1}}=-\mathbb{D}_{1},\quad {\cal
V}_{\Sigma_{c}^{*}D}^{X_{2.1}}=-\left(\mathbb{D}_{1}+\frac{1}{3}\mathbb{D}_{2}\bm{\sigma}\cdot\bm{T}\right)
\end{equation}
The isospin-isospin interaction and the $\mathcal{O}(\epsilon^2)$
contact terms are absorbed into $\mathbb{D}_1$ and $\mathbb{D}_2$.

\begin{figure*}[htp]
\centering
\includegraphics[width=0.41\textwidth]{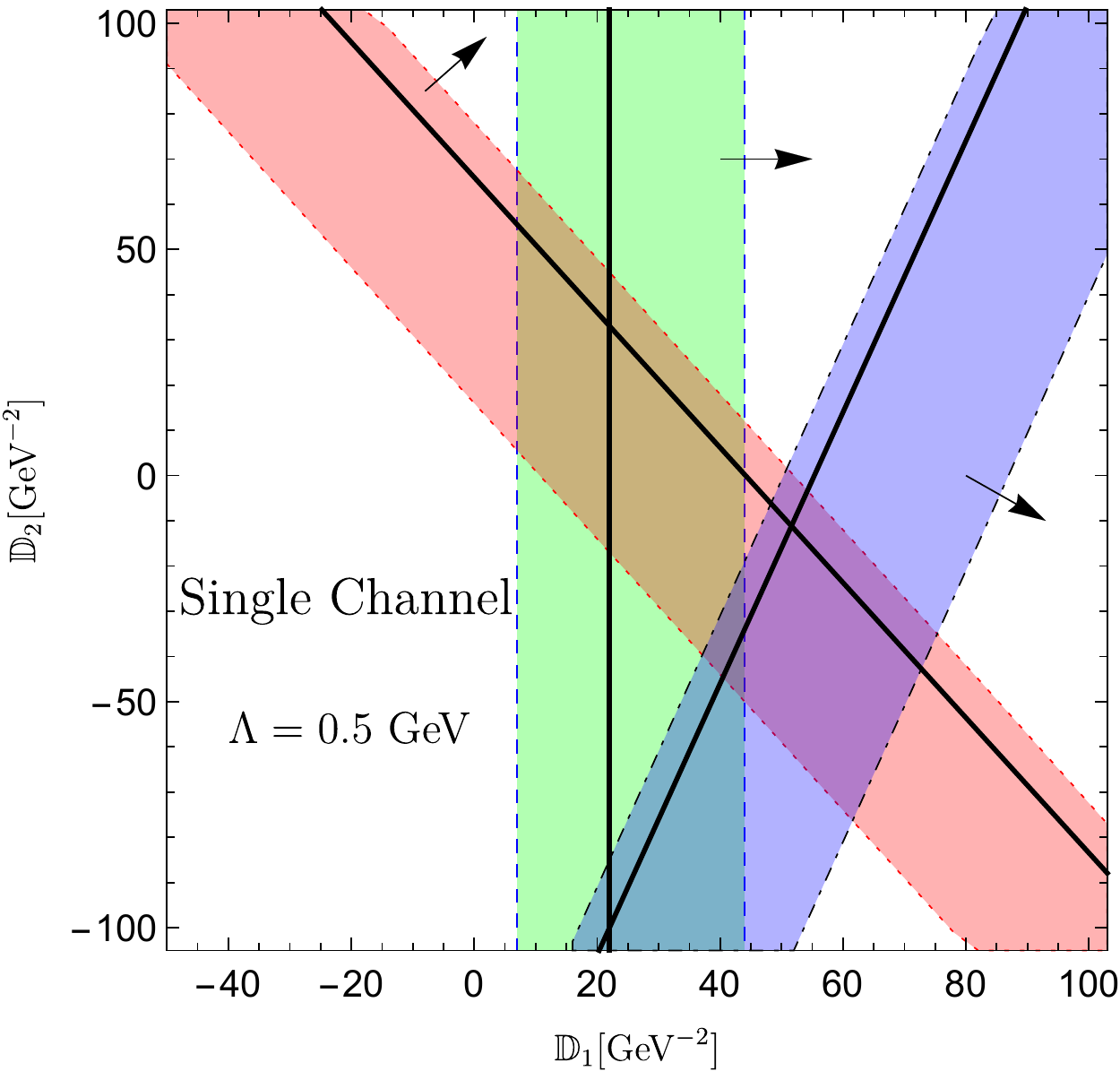}~~~~~
\includegraphics[width=0.4\textwidth]{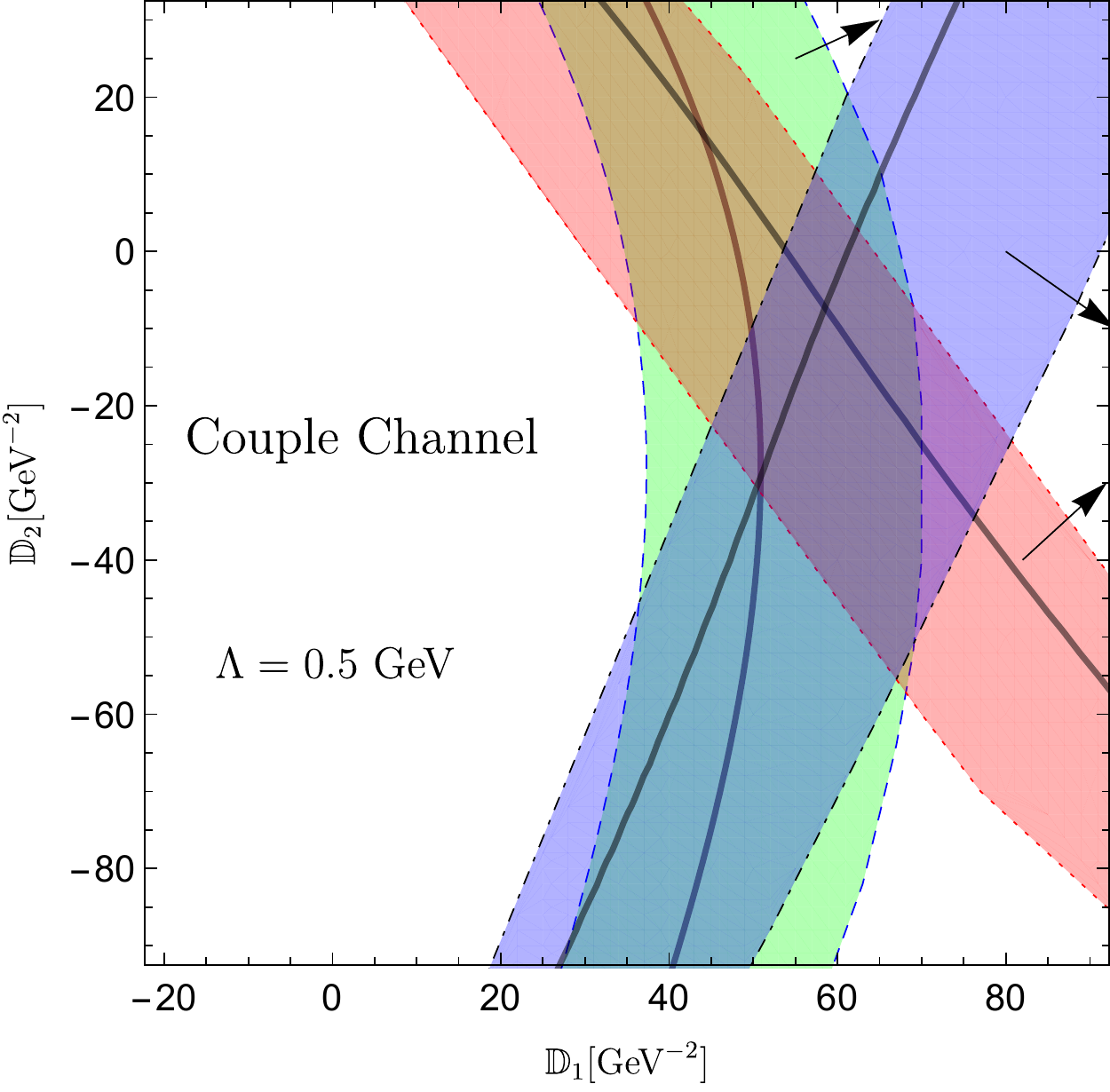}
\caption{Parameter regions of the contact terms in which there are
loosely bound states for $I={1\over 2}$ system. We choose
$\Lambda=0.5$ GeV. The regions of
$[\Sigma_c\bar{D}]^{I=1/2}_{J=1/2}$,
$[\Sigma_c\bar{D}^*]^{I=1/2}_{J=1/2}$ and
$[\Sigma_c\bar{D}^*]^{I=1/2}_{J=3/2}$ are surrounded by the dashed
line, dotted line and dot-dashed line, respectively. Every band
region corresponds to the binding energy $-30$ MeV$\sim0$ MeV. The
arrows give the directions that the bindings become deeper. The
solid lines are the sets of parameters corresponding to the three
$P_c$ states in Refs.~\cite{Aaij:2019vzc}. The left and right graphs
represent the results of scenarios II and III, respectively.
}\label{fig:search}
\end{figure*}

We vary $\mathbb{D}_1$ and $\mathbb{D}_2$ in the range from $-100
\text{ GeV}^{-2}$ to $100 \text{ GeV}^{-2}$, respectively. We show
the parameter regions in which there exist loosely bound states for
$[\Sigma_c\bar{D}]^{I=1/2}_{J=1/2}$,
$[\Sigma_c\bar{D}^*]^{I=1/2}_{J=1/2}$ and
$[\Sigma_c\bar{D}^*]^{I=1/2}_{J=3/2}$ in Fig.~\ref{fig:search},
where we choose $\Lambda=0.5$ GeV. Since the molecules are loosely
bound states, we adopt the binding energy $E=-30$ MeV as the lower
limit. In Fig.~\ref{fig:search}, there is a small region, in which
three bound states can coexist. In this region, the binding energy
ranges for $[\Sigma_c\bar{D}]^{I=1/2}_{J=1/2}$,
$[\Sigma_c\bar{D}^*]^{I=1/2}_{J=1/2}$ and
$[\Sigma_c\bar{D}^*]^{I=1/2}_{J=3/2}$ are $[-30,-25]$, $[-11,0]$ and
$[-8,-4]$ MeV, respectively.

\begin{figure*}[htp]
\centering
\begin{tabular}{ccc}
\includegraphics[width=0.32\textwidth]{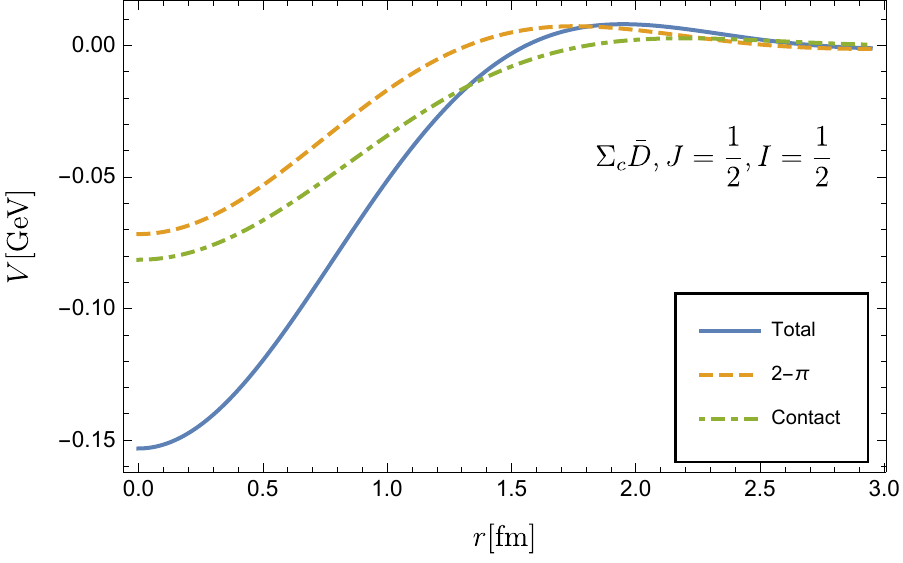}~&~
\includegraphics[width=0.32\textwidth]{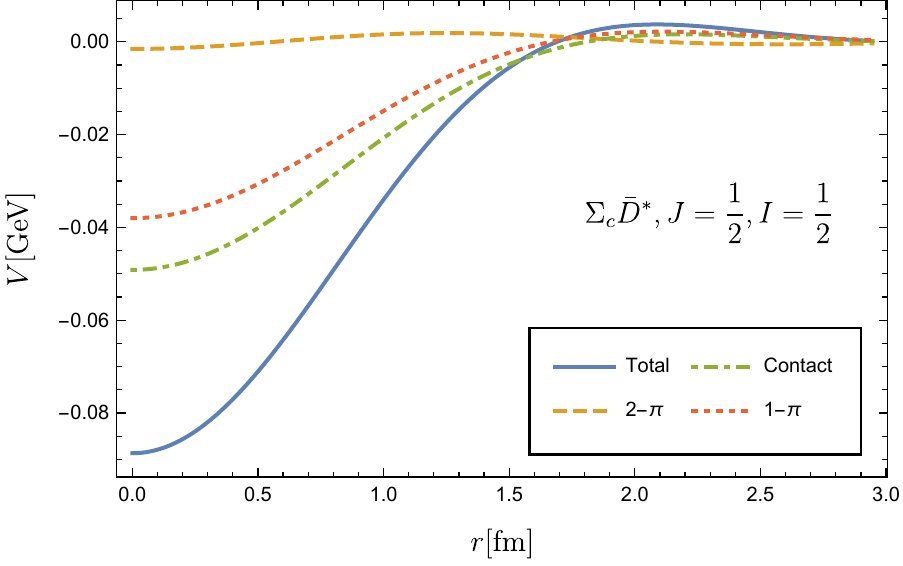}~&~
\includegraphics[width=0.32\textwidth]{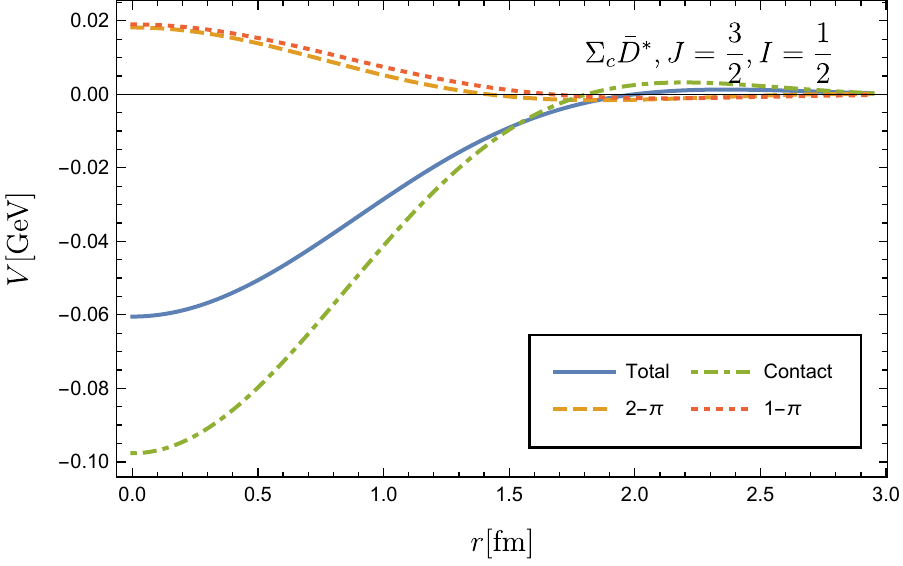}\\
\end{tabular}
\caption{The potentials for the $[\Sigma_c\bar{D}^{(*)}]^{I=1/2}$
systems in single channel calculation in scenario II. The cutoff
parameter $\Lambda=0.5$ GeV. The LECs of contact terms are
$\mathbb{D}_1= 42\text{ GeV}^{-2}$ and $\mathbb{D}_2= -25\text{
GeV}^{-2}$. }\label{fig:threebound}
\end{figure*}

We choose one set of parameters in the overlap region of three bands
in Fig.~\ref{fig:search}, $\mathbb{D}_1= 42\text{ GeV}^{-2}$ and
$\mathbb{D}_2= -25\text{ GeV}^{-2}$. The potentials are displayed in
Fig.~\ref{fig:threebound}, where the potentials for all three
channels are attractive. For the $[\Sigma_c\bar{D}]^{I=1/2}_{J=1/2}$
system, the potential from the contact terms and two-pion exchange
are both attractive. The binding energy is also deeper than that of
$P_c(4312)$ as a $\Sigma_{c}\bar{D}$ bound state. For the
$[\Sigma_c\bar{D}^*]^{I=1/2}_{J=1/2}$ system, the interaction of the
two-pion exchange are very weak. The attractive one-pion exchange
and contact interactions generate a loosely bound state. For the
$[\Sigma_c\bar{D}^*]^{I=1/2}_{J=3/2}$ system, both one-pion exchange
and two-pion exchange are repulsive. The loosely bound state arises
from the very attractive contact interaction. The bound state of
$[\Sigma_c\bar{D}^*]^{I=1/2}_{J=3/2}$ we got is dominated by the
short-distance contact interaction, which may arise from the vector
meson $\rho$ and $\omega$ exchange in the OBE mode.

We draw the binding energies of threes $P_c$ states as three solid
lines in Fig.~\ref{fig:search} if we assume they are molecular
states. There are three cross points in which two of three states
can coexist. The three cross points are not very close. In other
words, if we restrict the binding energies to the experimental
values, it is hard to reproduce the three states simultaneously.

\subsection{Scenario III}

In the above two scenarios, we only consider the potentials of the
elastic channels as shown in the upper panel of Fig.~\ref{fig:cc}.
We include the HQS partner states of the external lines as
intermediate states. Only part of the coupled-channel effects is
taken into account. For example, the contributions from the ladder
diagrams generated by the inelastic tree diagrams are dropped. In
scenario III, we improve our results by including the inelastic
channels and solving the coupled-channel Schr\"{o}dinger equation.

\begin{table}
    \caption{The channels we considered in the coupled-channel calculations. The bold ones are our channels of interest. }\label{tab:couplechannel}
\setlength{\tabcolsep}{4mm}
    \begin{tabular}{c|cccc}
        \toprule[1pt]\toprule[1pt]
        Channel & 1 &   2&  3 & 4 \tabularnewline
        \midrule[1pt]
        $J=\frac{1}{2}$ &   $\bm{\Sigma_{c}\bar{D}}$ &  $\bm{\Sigma_{c}\bar{D}^{*}}$    & $\Sigma_{c}^{*}\bar{D}^{*}$ & $\Sigma_{c}^{*}\bar{D}$  \tabularnewline
        $J=\frac{3}{2}$ &   $\bm{\Sigma_{c}\bar{D}^{*}}$ &  $\Sigma_{c}^{*}\bar{D}$ &   $\Sigma_{c}^{*}\bar{D}^{*}$ &    \tabularnewline
        \bottomrule[1pt]\bottomrule[1pt]
    \end{tabular}
\end{table}

\begin{table*}
    \caption{The numerical results in the coupled-channel calculations in scenario III. The $P_i$ is the proportion of the specific channel.}\label{tab:SIII}
\setlength{\tabcolsep}{6mm}
\begin{tabular}{cc|ccccc}
\toprule[1pt]\toprule[1pt]
    S-III & Exp.(MeV) & Mass$(\text{MeV})$ & RMS(\text{fm}) & $P_{1}(\%)$ & $P_{2}(\%)$ & $P_{3}(\%)$\tabularnewline
\midrule[1pt]
    $P_{c}(4312)$ & $4311.9\pm0.7_{-0.6}^{+6.8}$ & 4305 & 1.21 & 99.4 & 0.5 & 0.1\tabularnewline
    $P_{c}(4440)$ & $4440.3\pm1.3_{-4.7}^{+4.1}$ & 4446 & 1.22 & 1.0 & 98.0 & 0.9\tabularnewline
    $P_{c}(4457)$ & $4457.3\pm1.3_{-4.1}^{+0.6}$ & 4458 & 1.28 & 96.8 & 2.5 & 0.7\tabularnewline
\bottomrule[1pt]\bottomrule[1pt]
\end{tabular}
\end{table*}

For the $J={1\over 2}$ and $J={3\over 2}$ systems, four and three
channels can couple to one another, respectively, which are shown in
Table~\ref{tab:couplechannel}. The $[\Sigma_{c}\bar{D}]_{J=1/2}$,
$[\Sigma_c\bar{D}^*]_{J=1/2}$ and $[\Sigma_c\bar{D}^*]_{J=3/2}$ are
the channels we are interested in. For the other channels, we only
include their leading order potentials. We can get these potentials
either from the tree diagram calculations or the HQS analysis as
illustrated in Sec.~\ref{sec:hqs}. Both approaches lead to the same
results. The matrix elements of the $\bm{l}_1\cdot\bm{l}_2$ for
these channels read
\begin{eqnarray}
\langle\bm{l}_1\cdot\bm{l}_2\rangle_{J=1/2}&=&\left(\begin{array}{cccc}
0 & \frac{1}{\sqrt{3}} & -\frac{1}{\sqrt{6}} & 0\\
\frac{1}{\sqrt{3}} & -\frac{2}{3} & -\frac{1}{3\sqrt{2}} & 0\\
-\frac{1}{\sqrt{6}} & -\frac{1}{3\sqrt{2}} & -\frac{5}{6} & 0\\
0 & 0 & 0 & 0
\end{array}\right),\nonumber \\
\langle\bm{l}_1\cdot\bm{l}_2\rangle_{J=3/2}&=&\left(\begin{array}{ccc}
\frac{1}{3} & \frac{1}{2\sqrt{3}} & -\frac{\sqrt{5}}{6}\\
\frac{1}{2\sqrt{3}} & 0 & \sqrt{\frac{5}{12}}\\
-\frac{\sqrt{5}}{6} & \sqrt{\frac{5}{12}} & -\frac{1}{3}
\end{array}\right).
\end{eqnarray}
Since the off-diagonal terms in the Hamiltonian only arise from the
$\bm{l}_1\cdot\bm{l}_2$ interaction,  the forth channel of the
$J=1/2$ system does not couple with the other three channels in the
leading order potentials.

\begin{figure*}[htp]
    \centering
    \begin{tabular}{ccc}
        \includegraphics[width=0.32\textwidth]{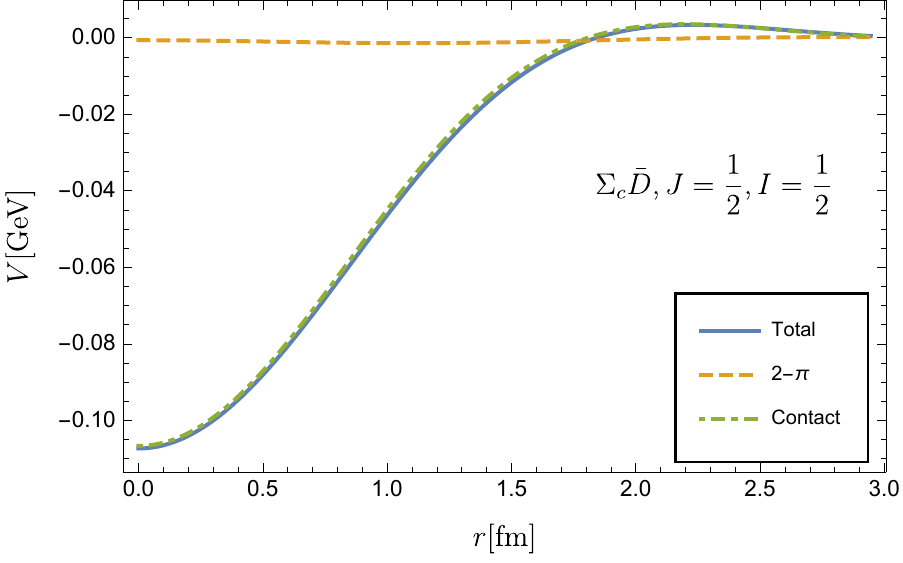}~&~
        \includegraphics[width=0.32\textwidth]{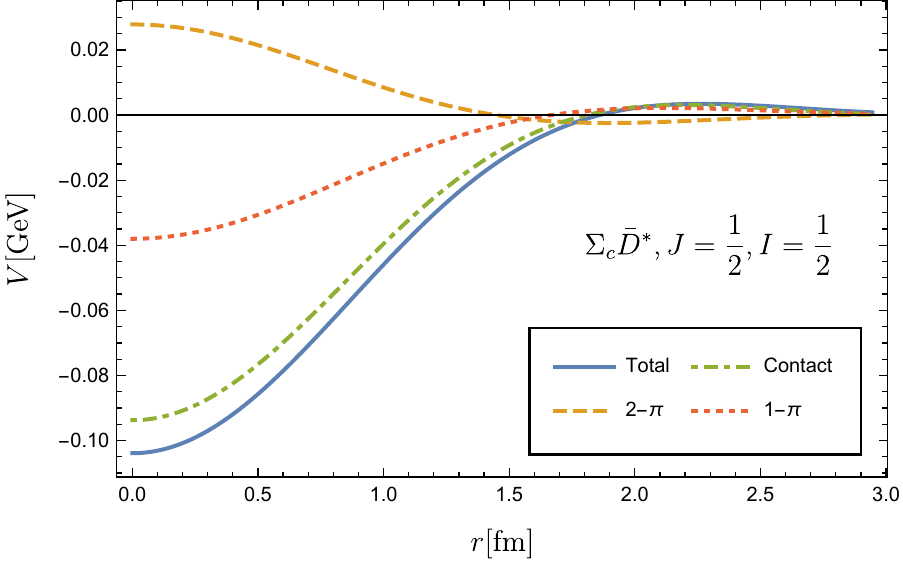}~&~
        \includegraphics[width=0.32\textwidth]{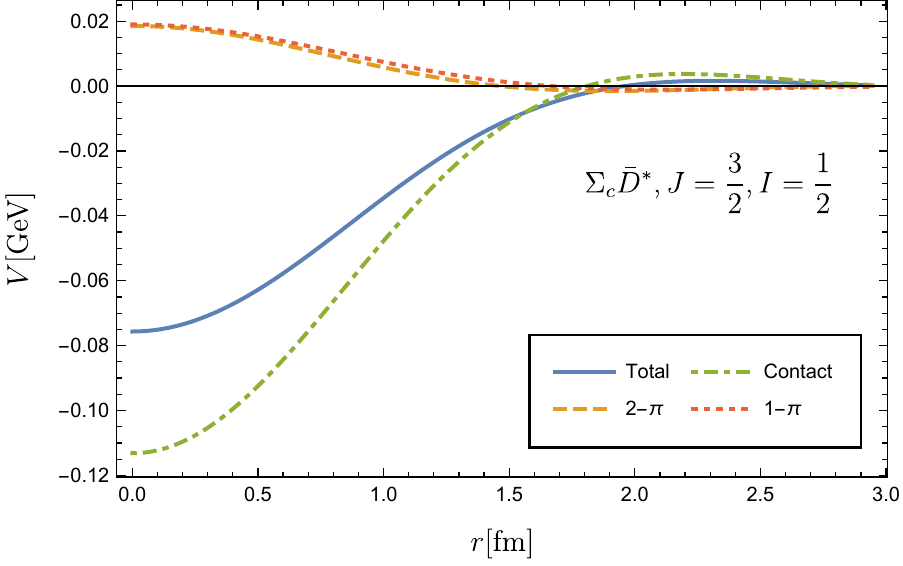}\\
    \end{tabular}
    \caption{The potentials for the $[\Sigma_c\bar{D}^{(*)}]^{I=1/2}$ systems in coupled-channel calculation in scenario III. We neglect the mass splittings in the box diagrams and remove the 2PIR contribution. The cutoff parameter $\Lambda=0.5$ GeV. The LECs of contact terms are $\mathbb{D}_1= 55\text{ GeV}^{-2}$ and $\mathbb{D}_2= -10\text{ GeV}^{-2}$. }\label{fig:ccpttl}
\end{figure*}

For our interested channels, we compute their potentials up to the
next-to-leading order. In the coupled-channel calculations, we deal
with the box diagrams in a different way. As illustrated in
Fig.~\ref{fig:cc}, we have included the inelastic tree diagrams at
the leading order. These diagrams will be iterated to generate the
ladder diagrams. Thus, to avoid the double counting of this
contribution, we ignore the mass splittings in the box diagrams and
remove the 2PR contributions.

We vary the two unknown LECs $\mathbb{D}_1$ and $\mathbb{D}_2$ as we
did in scenario II. The parameter regions in which the three loosely
bound states (with binding energy $-30$ MeV$\sim0$ MeV) can exist
are shown in the second graph of Fig.~\ref{fig:search}. Because of
the coupled-channel effect, the parameter regions are no longer
bands with fixed-width. We notice that the overlap region in which
three bound states can coexist is much larger than that in scenario
II. Meanwhile, the three cross points of the three lines
corresponding to $P_c(4312)$, $P_c(4440)$ and $P_c(4457)$ are much
closer than those in scenario II. Therefore, the coupled-channel
effect is very important to reproduce the three states
simultaneously.

We choose one set of LECs in the region that three molecular states
can coexist to give numerical results, where we take $\mathbb{D}_1=
55\text{ GeV}^{-2}$ and $\mathbb{D}_2= -10\text{ GeV}^{-2}$. The
masses, the RMSs and the proportion of each channel are given in
Table~\ref{tab:SIII}. With this set of LECs, we reproduce the three
$P_c$ states in the experiment as the molecular states
simultaneously. Among them, the coupled-channel effect is more
significant for the formation of $P_c(4457)$. We plot the potentials
of their dominant channels in Fig.~\ref{fig:ccpttl}. We notice that
the attractive interaction mainly stems from the contact
interaction, which is similar to that in scenario II. As we
discussed before, the dominant contact interaction may arise from
the vector meson $\rho$ and $\omega$ exchange in the OBE model.

\section{Discussion and Summary}\label{sec:sum}

In summary, we calculate the effective potentials of the
$\Sigma_c\bar{D}^{(*)}$ systems in the heavy hadron perturbation
theory. We adopt the small scale expansion to keep the mass
splittings between the HQS multiplets. We include the contact
interaction and one-pion exchange interaction at the leading order.
At the next-to-leading order, we take the two-pion exchange
interaction into consideration. The renormalizations of vertices and
wave functions are included by taking the physical values of the
parameters in chiral Lagrangians.

We employ the quark model with heavy quark spin symmetry to get some
relations between different systems. Our analytical results are
consistent with these relations in the heavy quark limit. Using
these relations, we obtain the potentials of the partner channels in
the HQS without calculating extra diagrams. We also show that the
HQS violation effect is not negligible in calculating the potentials
between the charmed hadrons. Since the molecular states are very
shallow bound states, their existence is very sensitive to the
potentials. The molecular states calculated in the charmed sector in
the heavy quark limit might be misleading. For the box diagrams,
taking the heavy quark limit will make the original two particle
irreducible diagrams reducible. Extra operation to remove the 2PR
contributions is needed. One should be cautious about the
uncertainty and trap of using the HQS to obtain the potentials
between the charmed hadrons.

Due to lack of experimental data, we can not determine the four LECs
in the contact terms precisely. We use three scenarios to estimate
the contact interactions. In the first scenario, we assume a contact
interaction at the quark level phenomenologically and then relate
the contact terms in nuclear force to those in the
$\Sigma_c\bar{D}^{(*)}$ systems. With the LECs, we reproduce the
$P_c(4312)$ and $P_c(4440)$ and predict three loosely bound states
in the $I={3\over 2}$ channels. We are unable to reproduce
$P_c(4457)$ due to the large uncertainty of LECs in the first
scenario. In the second scenario, we focus on the $I={1\over 2}$
channels. There are only two unknown independent LECs. We vary the
two LECs and search for the region that three $P_c$ states can
coexist as the loosely bound states. We do find a parameter region
in which we can reproduce three $P_c$ states simultaneously. The
region is very small. The solution corresponding to the $P_c(4457)$
seems slightly less natural, since the attractions all arise from
the short-range contact interactions. In the third scenario, we
consider the coupled-channel effect in the leading order on the
basis of scenario II. To avoid the double counting, we neglect the
mass splittings in the box diagrams and remove their 2PR
contributions. Through the coupled-channel calculations in scenario
III, we obtain a large parameter region in which three $P_c$ states
can coexist as molecular states. The attraction mainly comes from
the contact interactions.

We have reproduced the three $P_c$ states in our calculations. The
uncertainties come from either the framework or the LECs. In this
work, we only take the $S$-wave into consideration. However, the
$S$-$D$ wave mixing plays an important role in reproducing the
binding energy of the deutron. We do not consider the HQS violation
effect in the LECs. In order to reduce the number of LECs, we relate
them to each other through HQS. The approximation will introduce
errors, especially for the contact terms. The physical information
at the high energy scale is packaged into the contact terms, in
which the HQSS tends to be broken. Finally, the coupled-channel
effect can be considered more carefully. In this framework, we
calculate the potentials of  the interested channels to the
next-to-leading order. However, for the other inelastic channels, we
only calculate their leading order potentials. The numerical results
could be improved if one calculates all potentials to the
next-to-leading or even higher order.

The main uncertainty comes from the LECs of the contact
interactions. Thus the lattice QCD simulation on the
$\Sigma_c^{(*)}\bar{D}^{(*)}$ scattering is called for. Our
analytical results can be used to do chiral extrapolation for
lattice QCD. With the lattice QCD results in the coming future, the
LECs for the contact interaction can be determined more precisely.
Then the nature of the $P_c$ states in experiment can be identified
and more reliable predictions for the other systems can be given.

\section*{ACKNOWLEDGMENTS}

L. Meng is very grateful to X. Z. Weng, X. L. Chen and W. Z. Deng
for very helpful discussions. This project is supported by the
National Natural Science Foundation of China under Grants 11575008,
11621131001 and 973 program.

\begin{appendix}
\section{Matrix elements}\label{app:matrix}
The matrix elements of the isospin-isospin operator is
\begin{equation}
\langle \bm{I}_1\cdot\bm{I}_2\rangle= {1\over
2}[I(I+1)-I_1(I_1+1)-I_2(I_2+1)],
\end{equation}
where $I$, $I_1$ and $I_2$ are the total isospin, isospin of
$\Sigma_c$ and isospin of $\bar{D}^{(*)}$, respectively.

The $\bm{T}$ we defined is proportional to the spin operator
$\bm{S_2}$ of $\bar{D}^*$ as
\begin{equation}
\bm T=-\bm{S_2}.
\end{equation}
The matrix elements of spin-spin operator $\bm{S}_1\cdot\bm{S}_2$ is
\begin{equation}
\langle \bm{S}_1\cdot\bm{S}_2\rangle= {1\over
2}[J(J+1)-S_1(S_1+1)-S_2(S_2+1)],
\end{equation}
where $J$ is the total spin of $\Sigma_c^{(*)}\bar{D}^*$.

The matrix elements of the $\bm{l}_1\cdot\bm{l}_2$ spin-spin
operator in the light degrees of freedom can be calculated via the
spin rearrangement. Using the Wigner 9-J symbols, the
$\Sigma_c^{(*)}\bar{D}^{(*)}$ states can be related to the ones with
specific total light spin and heavy spin by the following relations,
\begin{eqnarray}
&~&|l_{1}h_{1}S_{1}l_{2}h_{2}S_{2}JM\rangle  \nonumber \\
    &=&\sum_{L,H}\sqrt{(2S_{1}+1)(2S_{2}+1)(2L+1)(2H+1)}\nonumber \\
&~&\times\left\{ \begin{array}{ccc}
l_{1} & l_{2} & L\\
h_{1} & h_{2} & H\\
S_{1} & S_{2} & J
\end{array}\right\} |l_{1}l_{2}Lh_{1}h_{2}HJM\rangle,
\end{eqnarray}
where $l_i$ and $h_i$ are the light spin and heavy spin for
$\Sigma_c^{(*)}$ or $\bar{D}^{(*)}$, respectively. $L$ and $H$ are
the total light spin and total heavy spin for the two particle
states, respectively. Thus, the matrix elements of the
$\bm{l}_1\cdot\bm{l}_2$ can be expressed as
\begin{eqnarray}
\langle \bm{l}_1\cdot\bm{l}_2\rangle &=&\langle l_{1}h_{1}S_{1}l_{2}h_{2}S_{2}JM|l_{1}\cdot l_{2}|l_{1}h_{1}S_{1}l_{2}h_{2}S_{2}JM\rangle\nonumber \\
&=&\sum_{L,H}\left[L(L+1)-l_{1}(l_{1}+1)-l_{2}(l{}_{2}+1)\right]\nonumber \\
&~&\times
\frac{1}{2}(2S_{1}+1)(2S_{2}+1)(2L+1)(2H+1)\nonumber \\
&~&\times \left\{ \begin{array}{ccc}
l_{1} & l_{2} & L\\
h_{1} & h_{2} & H\\
S_{1} & S_{2} & J
\end{array}\right\} ^{2} .
\end{eqnarray}

\section{Integrals}\label{app:integral}
\subsection{Definitions of integral functions}
We will use the ``M$x$B$y$'' to denote the integrals with $x$ light
meson propagators and $y$ heavy hadron propagators in the following.
\begin{widetext}
\begin{itemize}
\item M1B0
\begin{equation}
i\int\frac{d^{d}l\lambda^{4-d}}{(2\pi)^{d}}\frac{\{1,l^{\alpha},l^{\alpha}l^{\beta}\}}{l^{2}-m^{2}+i\varepsilon}\equiv\left\{
J_{0}^{c},0,g^{\alpha\beta}J_{21}^{c}\right\} (m),
\end{equation}
\item M2B0
\begin{equation}
i\int\frac{d^{d}l\lambda^{4-d}}{(2\pi)^{d}}\frac{\{1,l^{\alpha},l^{\alpha}l^{\beta},l^{\alpha}l^{\beta}l^{\gamma}\}}{\left(l^{2}-m^{2}+i\varepsilon\right)\left[(l+q)^{2}-m^{2}+i\varepsilon\right]}\equiv\left\{
J_{0}^{F},q^{\alpha}J_{11}^{F},q^{\alpha}q^{\beta}J_{21}^{F}+g^{\alpha\beta}J_{22}^{F},(g\vee
q)J_{31}^{F}+q^{\alpha}q^{\beta}q^{\gamma}J_{32}^{F}\right\} (m,q),
\end{equation}
\item M1B1
\begin{equation}
i\int\frac{d^{d}l\lambda^{4-d}}{(2\pi)^{d}}\frac{\{1,l^{\alpha},l^{\alpha}l^{\beta},l^{\alpha}l^{\beta}l^{\gamma}\}}{\left(v\cdot
l+\omega+i\varepsilon\right)\left(l^{2}-m^{2}+i\varepsilon\right)}\equiv\left\{
J_{0}^{a},v^{\alpha}J_{11}^{a},v^{\alpha}v^{\beta}J_{21}^{a}+g^{\alpha\beta}J_{22}^{a},(g\vee
v)J_{31}^{a}+v^{\alpha}v^{\beta}v^{\gamma}J_{32}^{a}\right\}
(m,\omega),
\end{equation}
\item M2B1
\begin{eqnarray}
    &&i\int\frac{d^{d}l\lambda^{4-d}}{(2\pi)^{d}}\frac{\{1,l^{\alpha},l^{\alpha}l^{\beta},l^{\alpha}l^{\beta}l^{\gamma},l^{\alpha}l^{\beta}l^{\gamma}l^{\delta}\}}{\left(v\cdot l+\omega+i\varepsilon\right)\left(l^{2}-m^{2}+i\varepsilon\right)\left[(l+q)^{2}-m^{2}+i\varepsilon\right]}\equiv\Big\{ J_{0}^{T},q^{\alpha}J_{11}^{T}+v^{\alpha}J_{12}^{T},g^{\alpha\beta}J_{21}^{T}+q^{\alpha}q^{\beta}J_{22}^{T}+v^{\alpha}v^{\beta}J_{23}^{T}\nonumber \\
&&  \quad+(q\vee v)J_{24}^{T},(g\vee q)J_{31}^{T}+q^{\alpha}q^{\beta}q^{\gamma}J_{32}^{T}+(q^{2}\vee v)J_{33}^{T}+(g\vee v)J_{34}^{T}+(q\vee v^{2})J_{35}^{T}+v^{\alpha}v^{\beta}v^{\gamma}J_{36}^{T},(g\vee g)J_{41}^{T} \nonumber \\
    &&\quad+(g\vee q^{2})J_{42}^{T}+q^{\alpha}q^{\beta}q^{\gamma}q^{\delta}J_{43}^{T}+(g\vee v^{2})J_{44}^{T}+v^{\alpha}v^{\beta}v^{\gamma}v^{\delta}J_{45}^{T}+(q^{3}\vee v)J_{46}^{T}+(q^{2}\vee v^{2})J_{47}^{T}+(q\vee v^{3})J_{48}^{T} \nonumber \\
&&  \quad+(g\vee q\vee v)J_{49}^{T}\Big\}(m,\omega,q),
\end{eqnarray}
\item M2B2
\begin{eqnarray}
    &&i\int\frac{d^{d}l\lambda^{4-d}}{(2\pi)^{d}}\frac{\{1,l^{\alpha},l^{\alpha}l^{\beta},l^{\alpha}l^{\beta}l^{\gamma},l^{\alpha}l^{\beta}l^{\gamma}l^{\delta}\}}{\left(v\cdot l+\omega_{1}+i\varepsilon\right)\left[(+/-)v\cdot l+\omega_{2}+i\varepsilon\right]\left(l^{2}-m^{2}+i\varepsilon\right)\left[(l+q)^{2}-m^{2}+i\varepsilon\right]}\equiv\Big\{ J_{0}^{R/B},q^{\alpha}J_{11}^{R/B}+v^{\alpha}J_{12}^{R/B},\nonumber\\
    &&\quad g^{\alpha\beta}J_{21}^{R/B} +q^{\alpha}q^{\beta}J_{22}^{R/B}+v^{\alpha}v^{\beta}J_{23}^{R/B}+(q\vee v)J_{24}^{R/B},(g\vee q)J_{31}^{R/B}+q^{\alpha}q^{\beta}q^{\gamma}J_{32}^{R/B}+(q^{2}\vee v)J_{33}^{R/B}+(g\vee v)J_{34}^{R/B}\nonumber\\
    &&  \quad+(q\vee v^{2})J_{35}^{R/B}+v^{\alpha}v^{\beta}v^{\gamma}J_{36}^{R/B},(g\vee g)J_{41}^{R/B}+(g\vee q^{2})J_{42}^{R/B}+q^{\alpha}q^{\beta}q^{\gamma}q^{\delta}J_{43}^{R/B}+(g\vee v^{2})J_{44}^{R/B}+v^{\alpha}v^{\beta}v^{\gamma}v^{\delta}J_{45}^{R/B}\nonumber\\
    &&\quad +(q^{3}\vee v)J_{46}^{R/B}+(q^{2}\vee v^{2})J_{47}^{R/B}+(q\vee v^{3})J_{48}^{R/B}+(g\vee q\vee v)J_{49}^{R/B}\Big\}(m,\omega_{1},\omega_{2},q),
\end{eqnarray}
\end{itemize}
We use representation $X\vee Y \vee Z...$ to denote the symmetrized
tensor structure for concise. For example, 
\begin{eqnarray}
q\vee v &\equiv&
q^{\alpha}v^{\beta}+q^{\beta}v^{\alpha},\nonumber\\
g\vee q &\equiv&
g^{\alpha\beta}q^{\gamma}+g^{\alpha\gamma}q^{\beta}+g^{\gamma\beta}q^{\alpha},\nonumber\\
g\vee g&\equiv& g^{\alpha\beta}g^{\gamma
	\delta}+g^{\alpha\gamma}g^{\beta\delta}+g^{\alpha\delta}g^{\beta\gamma}.
\end{eqnarray} 
\end{widetext}

\subsection{Calculations of integral functions}

In Ref.~\cite{Wang:2018atz}, the authors calculated these loop
integrals directly. While in this work, we calculate these loop
integrals through a different way. We take the $J^a_x$ as an
example. We calculate the $J^a_0$ directly as in
Ref.~\cite{Wang:2018atz}. Then, we use $v_{\alpha}$ to contract with
the $J^a_{1x}$ terms, i.e.,
\begin{eqnarray}
J_{11}^{a}v^{2} &=&i\int\frac{d^{d}l\lambda^{4-d}}{(2\pi)^{d}}\frac{v\cdot l}{\left(v\cdot l+\omega+i\varepsilon\right)\left(l^{2}-m^{2}+i\varepsilon\right)}\nonumber \\
    &=&i\int\frac{d^{d}l\lambda^{4-d}}{(2\pi)^{d}}\frac{1}{\left(l^{2}-m^{2}+i\varepsilon\right)}\left[1-\frac{\omega}{\left(v\cdot l+\omega+i\varepsilon\right)}\right]\nonumber \\
    &=&J_{0}^{c}-\omega J_{0}^{a},
\end{eqnarray}
Here, we get the relation of $J_{11}^{a}$ with the known functions.
Third, we can use $g_{\alpha\beta}$ and $v_{\alpha}$ to contact with
the $J_{2x}^a$ terms,
\begin{eqnarray}
&&J_{21}^{a}v^{2}+dJ_{22}^{a} \nonumber \\
&=&i\int\frac{d^{d}l\lambda^{4-d}}{(2\pi)^{d}}\frac{l^{2}}{\left(v\cdot l+\omega+i\varepsilon\right)\left(l^{2}-m^{2}+i\varepsilon\right)}\nonumber \\
    &=&i\int\frac{d^{d}l\lambda^{4-d}}{(2\pi)^{d}}\frac{1}{\left(v\cdot l+\omega+i\varepsilon\right)}\left[1+\frac{m^{2}}{\left(l^{2}-m^{2}+i\varepsilon\right)}\right]\nonumber \\
    &=&m^{2}J_{0}^{c},
    \end{eqnarray}
    \begin{eqnarray}
&&J_{21}^{a}v_{\beta}+J_{22}^{a}v_{\beta} \nonumber \\
    &=&i\int\frac{d^{d}l\lambda^{4-d}}{(2\pi)^{d}}\frac{v\cdot ll_{\beta}}{\left(v\cdot l+\omega+i\varepsilon\right)\left(l^{2}-m^{2}+i\varepsilon\right)}\nonumber \\
    &~~&\text{...}.
\end{eqnarray}
Therefore, following these procedures, we can relate the complicated
integrals to the simple ones step by step. The final results we get
are equivalent to those in Ref.~\cite{Wang:2018atz}.

In the following, we give the results of some simple integrals
first. The complicated ones can be reexpressed with them,
\begin{eqnarray}
J_{0}^{c}(m,q)&=&\frac{m^{2}}{16\pi^{2}}\ln\frac{m^{2}}{\lambda^{2}}+2m^{2}L,\\
J_{0}^{F}(m,q)&=&\int_{0}^{1}dz\frac{1}{16\pi^{2}}\left(1+\text{ln}\frac{\bar{\Delta}}{\lambda^{2}}\right)+2L(\lambda),
\end{eqnarray}
where $\bar{\Delta}(z)=m^{2}+q^{2}(z-1)z-i\varepsilon$. When
$q^{2}<0$, one get
\begin{equation}
J_{0}^{F} (m,q)
=-\frac{1}{16\pi^{2}}\left(1-\text{ln}\frac{m^{2}}{\lambda^{2}}-r\text{ln}|\frac{1+r}{1-r}|\right)+2L,
\end{equation}
where $r=\sqrt{|1-\frac{4m^{2}}{q^{2}}|}$.
\begin{eqnarray}
&&J_{0}^{a}(m,\omega,q) =\frac{1}{8\pi}\sqrt{m^{2}-\omega^{2}-i\varepsilon}\nonumber \\
    &&\quad+\int_{-\omega}^{0}dy\frac{2}{16\pi^{2}}\left(1+\text{ln}\frac{\tilde{\Delta}(y)}{\lambda^{2}}\right)+4\omega L,
\end{eqnarray}
where $\tilde{\Delta}(y)=m^{2}+y^{2}-\omega^{2}-i\varepsilon$. The
$J^{T}_x$ used in this work reads,
\begin{widetext}
\begin{eqnarray}
&&J^T_{21}(m,\omega,q)=\frac{2 (J^a_0+2 J^F_0 \omega )+J^T_0 \left(4 m^2-q^2-4 \omega ^2\right)}{4 (d-2)},\\
&&J^T_{31}(m,\omega,q)=\frac{J^T_0 \left(-4 m^2+q^2+4 \omega ^2\right)-2 (J^a_0+2 J^F_0 \omega )}{8 (d-2)},\\
&&J^T_{32}(m,\omega,q)=\frac{6 ((3-d) J^a_0+2 J^F_0 \omega )+J^T_0 \left(-(d+1) q^2+12 m^2-12 \omega ^2\right)}{8 (d-2) q^2},\\
&&J^T_{33}(m,\omega,q)=\frac{1}{4 (d-2) (d-1) q^2}\Big[-2 \left(d^2-4 d+3\right) J^a_0 \omega +2 d^2 J^c_0+d^2 J^F_0 q^2-8 d J^c_0-4 d J^F_0 m^2-2 d J^F_0 q^2\nonumber \\
 &&~~~~~~~~~~~~~~~~~~ +4 d J^F_0 \omega ^2-(d-1) J^T_0 \omega  \left((d-1) q^2-4 m^2+4 \omega ^2\right)+8 J^c_0+8 J^F_0 m^2-4 J^F_0 \omega ^2\Big],\\
&&J^T_{34}(m,\omega,q)=\frac{1}{4 (d-2) (d-1)}\Big[-2 (d-1) J^a_0 \omega +2 d J^c_0+4 d J^F_0 m^2-d J^F_0 q^2-4 d J^F_0 \omega ^2+(d-1) J^T_0 \omega  \left(-4 m^2+q^2+4 \omega ^2\right)\nonumber \\
&&~~~~~~~~~~~~~~~~~~ -4 J^c_0-8 J^F_0 m^2+2 J^F_0 q^2+4 J^F_0 \omega ^2\Big],\\
&&J^T_{36}(m,\omega,q)=\frac{1}{4 (d-2) (d-1)}\Big[2 \left(J^F_0 \left(2 \left(d^2-1\right) \omega ^2-4 (d-2) m^2+(d-2) q^2\right)+3 (d-1) J^a_0 \omega -2 (d-2) J^c_0\right)\nonumber \\
 &&~~~~~~~~~~~~~~~~~~ +(1-d) J^T_0 \omega  \left(4 (d+1) \omega ^2-12 m^2+3 q^2\right)\Big],\\
&&J^T_{41}(m,\omega,q)=\frac{1}{16 (d-2) (d-1) d}\Big[2 (J^a_0 (4 (2 d-3) m^2-d q^2+4 (3-2 d) \omega ^2+q^2)+2 \omega  (4 (d-2) J^c_0+J^F_0 (4 (2 d-3) m^2\nonumber \\
&&~~~~~~~~~~~~~~~~~~ +(3-2 d) q^2-4 (d-1) \omega ^2)))+(d-1) J^T_0 (-4 m^2+q^2+4 \omega ^2)^2\Big],\\
&&J^T_{42}(m,\omega,q)= \frac{1}{16 (d-2) (d-1) d q^2}\Big[2 (2 \omega  (J^F_0 ((d^2+d-3) q^2+(12-8 d) m^2+4 (d-1) \omega ^2)+2 (d-2)^2 J^c_0)\nonumber \\
 &&~~~~~~~~  +(d-1) J^a_0 (4 (d-3) m^2+(d+1) q^2-4 (d-3) \omega ^2))+(d-1) J^T_0 (4 m^2-q^2-4 \omega ^2) (d q^2-4 m^2+q^2+4 \omega ^2)\Big],\\
&&J^T_{43}(m,\omega,q)=\frac{1}{16 (d-2) (d-1) d q^4}\Big[2 ((d^2-4 d+3) J^a_0 (7 d q^2-12 m^2+q^2+12 \omega ^2)-6 \omega  (J^F_0 ((2 d^2-3) q^2+(12-8 d) m^2 \nonumber \\
&&~~~~~~~~~~~~~~~~~~ +4 (d-1) \omega ^2)+2 (d-2)^2 J^c_0))+(d-1) J^T_0 ((d^2+4 d+3) q^4-24 m^2 (d q^2+q^2+4 \omega ^2) \nonumber \\
&&~~~~~~~~~~~~~~~~~~ +24 (d+1) q^2 \omega ^2+48 m^4+48 \omega ^4)\Big],\\
&&J_{x}^{R}(m,\omega_{1},\omega_{2},q)=\begin{cases}
-\frac{1}{\omega_{1}-\omega_{2}}\left[J_{x}^{T}(m,\omega_{1},q)-J_{x}^{T}(m,\omega_{2},q)\right] & \text{if }\omega_{1}\neq\omega_{2}  \\
-\frac{\partial}{\partial\omega}J_{x}^{T}(m,\omega,q)|_{\omega\rightarrow\omega_{1}(\text{or
}\omega_{2})} & \text{if }\omega_{1}=\omega_{2}
\end{cases}
\end{eqnarray}
\end{widetext}
The results of $J^B_x(m,\omega_1,\omega_2,q)$ depend on whether we
remove the 2PR contributions or not. When $\omega_1=-\omega_2$ there
is the pinch singularity. We use the following procedures to remove
it,
\begin{eqnarray}
    &~~&\int\frac{dl^{0}}{(2\pi)}\frac{f(l^{0},\bm{l})}{\left(v\cdot l+\omega_{1}+i\varepsilon\right)\left[-v\cdot l-\omega_{1}+i\varepsilon\right]}\nonumber \\
    &=&\int\frac{dl^{0}}{(2\pi)}\frac{-f(l^{0},\bm{l})}{\left(v\cdot l+\omega_{1}\right)^{2}},
\end{eqnarray}
where $f(l^0,\bm{l})$ is the other part of the $J^{B}_x$ integrals.
In the derivations, the principal integral is used, i.e.,
\begin{eqnarray}
\lim_{\varepsilon\rightarrow0^{+}}\frac{1}{x\pm
i\varepsilon}=\mathcal{P}\frac{1}{x}\mp i\pi\delta(x).
\end{eqnarray}
This procedure is equivalent to removing the contributions from the
poles of the matter fields. When $\omega_1\neq-\omega_2$, we
calculate the $J^B_x(m,\omega_1,\omega_2,q)$ as before. The results
read
\begin{eqnarray}
J^{B}_{nx}=\frac{1}{\omega_{1}+\omega_{2}}[J_{nx}^{T}(\omega_{1})+J_{nx}^{T}(\omega_{2})(-1)^{n+n_q}],
\end{eqnarray}
where $n$ and $n_q$ are the numbers of Lorentz indices and momentum
$q$ in the Lorentz structures. $n+n_q$ of all the integrals involved
in this work are even.

\end{appendix}

\vfil \thispagestyle{empty}

\newpage
\bibliography{Bib}

\end{document}